\documentclass[preprintnumbers,nofootinbib,amsmath,amssymb]{revtex4}

\def\ga{\mathrel{\raise.3ex\hbox{$>$\kern-.75em\lower1ex\hbox{$\sim$}}}}
\def\la{\mathrel{\raise.3ex\hbox{$<$\kern-.75em\lower1ex\hbox{$\sim$}}}}

\newcommand\beq{\begin{equation}}
\newcommand\eeq{\end{equation}}
\newcommand\beqar{\begin{eqnarray}}
\newcommand\eeqar{\end{eqnarray}}
\usepackage{graphicx}
\usepackage{epsfig}
\usepackage{epsf}
\usepackage{rotating}
\usepackage{verbatim}

\usepackage{multirow}

\begin{document}

\title{Spectrum of Perturbations in Anisotropic Inflationary Universe with Vector Hair}

\author{Burak Himmetoglu}
\affiliation{{\it School of Physics and Astronomy, University of Minnesota, Minneapolis, MN 55455, USA}}

\begin{abstract}
We study both the background evolution and cosmological
perturbations of anisotropic inflationary models supported by
coupled scalar and vector fields. The models we study preserve the $U(1)$
gauge symmetry associated with the vector field, and therefore do not possess instabilities 
associated with longitudinal modes (which instead plague some recently proposed models of
vector inflation and curvaton). We first intoduce a model in which the background anisotropy 
slowly decreases during inflation; we then confirm the stability of the background solution 
by studying the quadratic action for all the perturbations of the model. We then compute 
the spectrum of the $h_{\times}$ gravitational wave polarization. The spectrum we find
breaks statistical isotropy at the largest scales and reduces to the standard nearly scale 
invariant form at small scales. We finally discuss the possible relevance of our results to the
large scale CMB anomalies. 
\end{abstract}

\preprint{UMN-TH-2821/09}
\preprint{ArXiv:0910.3235}

\maketitle

\section{Introduction}

The inflationary paradigm has become a widely accepted description
of the early universe, which has been successful in solving the
classical problems of FRW cosmology~\cite{Guth,Linde}. Moreover,
inflation provides a mechanism for generation of nearly scale
invariant spectrum of perturbations that can seed the structure we
observe today~\cite{mfb}. Inflationary era is characterized by a
quasi de-Sitter expansion with a nearly constant Hubble rate.
During inflation, quantum fluctuations are generated and amplified
to scales above the Hubble scale (which specifies the horizon of
casual processes) where they remain frozen until they re-enter the
horizon after inflation has ended. A key assumption about the
quantum fluctuations during the quasi de-Sitter stage is that they
are in an adiabatic vacuum stage in the deep sub-horizon (early
time) regime, which results in a nearly scale invariant spectrum.

Results of the WMAP experiment~\cite{WMAP} are in overall
agreement with the predictions of inflation. However, certain
features of the full sky CMB maps seem to be anomalous in the
standard picture. These anomalies include the low power in the
quadrupole moment~\cite{lowl}, the alignment of the lowest
multipole moments (known as the 'axis-of-evil')~\cite{axis}, and
an asymmetry in power between the northern and southern ecliptic
hemispheres~\cite{asym}. The statistical significance of these
anomalies has been extensively discussed in the literature and
despite numerous efforts, an explanation due to a systematic
effect or a foreground signal affecting the analysis is not
forthcoming. These anomalies suggest a violation of statistical
isotropy of cosmological fluctuations, which is at odds with
standard mechanisms of generation of perturbations in inflationary
models. Therefore, the possibility of relating the anomalies with
modifications in the standard inflationary picture has been
considered by numerous authors recently.

In the standard picture, $a_{lm}$ coefficients of CMB temperature
anisotropies satisfy $< a_{lm}\, a_{lm}^* > = C_l\, \delta_{ll'}\,
\delta_{mm'}$ (so, the temperature fluctuations are said to be
statistically isotropic). It has been shown in~\cite{gcp,gcp2}
that an anisotropic expansion that took place at the onset of
standard single field inflation, leads to a nonvanishing
correlation of the $a_{lm}$ coefficients for different $l$-modes.
This violation of statistical isotropy might then be related to
the alignment of the lowest multipoles observed in the CMB data.
Therefore, several authors considered inflationary models with an
anisotropic stage. As the simplest possible modification,
anisotropic initial conditions at the onset of single field
inflation have been considered in
references~\cite{gcp,gcp2,uzan1,uzan2}. Isotropization of the
universe takes place due to the fast expansion supported by the
inflaton field. Fluctuations at scales comparable to the horizon 
scale during the anisotropic stage of inflation are sensitive 
to the background evolution and thus provide breaking of statistical
isotropy at those scales. If inflation had a limited duration, this 
could leave an imprint on the largest observable scales. On the other hand, fluctuations that leave
the horizon after isotropization re-enter earlier than modes that left
the horizon during the anisotropic stage, producing the standard nearly
scale invariant spectrum at small scales~\footnote{It was also
shown in~\cite{gkp} that the anisotropic stage 
could lead to a detectable gravitational wave signal.}. Since in such models isotropy is reached very soon
(within one Hubble time) the breaking of statistical isotropy can be visible 
if the total duration of inflation is tuned to only the minimal duration
to solve the standard problems of FRW cosmology.  

The fine-tuning can be relaxed if the anisotropic stage is
prolonged, due to the existence of other sources, for example vector fields. Recently,
anisotropic expansion driven by vector fields have been studied by
many authors. The anisotropic expansion is achieved when the
vector field acquires a nonzero vacuum expectation value (VEV)
along a spatial direction. There is a range of models which differ
by the way the VEV is obtained. The first anisotropic inflationary
model was considered long ago in reference~\cite{ford}, where the
$U(1)$ symmetry of the vector field action is broken by a
potential term, which needs to have a negative curvature to
support an anisotropic expansion. It has been shown in
reference~\cite{hcp1} that this model is plagued by a ghost
instability, and its quantum theory is inconsistent~\footnote{One
may hope that this model has a well behaved UV completion.
However, inflationary predictions are sensitive to this high
energy regime, where the completion is needed.}. In another model,
the ACW model~\cite{acw}, the VEV is obtained by a lagrange
multiplier field which fixes the norm of the vector field and
breaks the $U(1)$ gauge symmetry. The stability analysis of the
ACW model by considering the most general perturbations of the
background has been performed in references~\cite{dgw}
and~\cite{hcp1,hcp2}. In the latter studies, it has been shown
that the linearized perturbations diverge close to horizon
crossing, indicating the instability of the background solution.
Another model~\cite{soda1} considered a nonminimal coupling of the
vector field to the scalar curvature which breaks the conformal 
invariance of the vector field action. The coupling with curvature allows
for a slow-roll phase during the prolonged anisotropic expansion, and the 
universe isotropizes due to the existence of a massive scalar
field, which is responsible for the overall isotropic expansion of
the universe. Models of inflation with nonmininal coupling to
curvature have also been considered for standard isotropic
inflation~\cite{mukhvect}. Isotropy of space is realized if $3N$
mutually orthogonal vector fields have equal VEVs. Instead, for
randomly oriented $N$ vector fields, one expects a statistically
isotropic background with order $1/\sqrt{N}$ anisotropy. 
Perturbative calculations based around such a background
configuration have been considered in
references~\cite{Golovnev:2008hv, Golovnev:2009ks, Golovnev:2009rm} and in~\cite{hcp3}.
The latter study is the only complete study linearized study of 
perturbations, taking into account all the physical degrees of freedom, 
and coupling between vector field and metric perturbations. It has been shown
in~\cite{hcp3} that the equations of motion for the linearized perturbations
become singular close to horizon crossing, leading to instabilities. In all of these 
models (including~\cite{acw,soda1,mukhvect}), instabilities are related to the
existence of the longitudinal polarization of the vector field (which would otherwise 
be absent when the $U(1)$ symmetry is restored), which becomes a ghost close to horizon
crossing (In the vector inflation model~\cite{mukhvect}, ghosts also appear in the deep
UV regime). 

More recently, reference~\cite{Watanabe:2009ct} introduced an
anisotropic inflation model driven simultaneously by a vector
field and a massive scalar field. The vector field is massless,
but it is coupled to the scalar field through its kinetic term.
Such type of coupling preserves the $U(1)$ gauge invariance of the
vector field, so that the dangerous longitudinal vector
polarization is absent; moreover the conformal invariance of the
vector field is broken. When the vector field has a nonvanishing
VEV along a spatial direction, the model possess an
anisotropically expanding attractor solution. The anisotropy in
the attractor is initially small, but it increases towards the end
of inflation; therefore this is a counter example to the cosmic no
hair conjecture (See~\cite{Kaloper:1991rw} for a different
example) .The breaking of conformal invariance due to the
scalar-vector coupling can also be used to generate magnetic
fields from inflation, as discussed in
reference~\cite{Turner:1987bw} and more recently
in~\cite{Gordon:2000hv, Demozzi:2009fu} and~\cite{Watanabe:2009ct}. These models are expected 
to be free from ghost instabilities as long as the coupling to the scalar
field remains positive. However to our knowledge, there has been no complete
study of the stability of these models, which take into account all the physical
degrees of freedom of the system (including gravity). A complete study of stability is therefore necessary,
given the problems identified with other vector field models~\cite{hcp1,hcp2,hcp3}.

In this work, we will present a complete study of cosmological
perturbations of models with scalar-vector coupling that lead to
an anisotropically expanding Bianchi-I background solution. We
will also consider generalizations of the original model
introduced in~\cite{Watanabe:2009ct}, to include additional scalar
fields, so that isotropization can be achieved before the end of
inflation. This is required in order to obtain a scale invariant
spectrum of perturbations at small scales, but the large scale
spectrum is modified, which in turn can be related to the CMB
anomalies. Our study has two main steps: one is to show 
that the model is free of ghost instabilities, and the second is to study the 
resulting phenomenology. We perform the first step explicitly in this paper.
We develop the necessary tools to study the phenomenology of the model and 
as a simple exercise, we study only the spectrum of $h_{\times}$ gravitational
wave polarization in this paper.  The study of the full phenomenology (including the
spectrum of the curvature perturbation) will be communicated elsewhere~\cite{me}.

We will follow the formalism developed in~\cite{gcp2,hcp2}
to decompose and classify the perturbations. The generic
background metric we study is the Bianchi-I metric with a residual
two dimensional isotropy, given by
\begin{equation}
ds^2 = -dt^2 + a(t)^2\, dx^2 + b(t)^2\, \left( dy^2 + dz^2 \right)
\nonumber
\end{equation}
We exploit the two dimensional isotropy of the $y-z$ plane to
decompose the perturbations into two decoupled classes: $2d$
scalar modes and $2d$ vector modes (As discussed in~\cite{gcp2},
there is no two dimensional transverse-traceless mode). The
linearized Einstein equations and the quadratic action for the two
types of modes are decoupled and therefore we study them
separately. We can deduce the number of degrees of freedom coming
from each of the fields (gravity+vector field+scalar field) by a
simple counting, which does not depend on the decomposition chosen
to classify them. The metric has $10$ perturbations ($\delta
g_{\mu\nu}$) to start with, 4 of which can be removed by
coordinate transformations. Of the remaining 6 modes, 4 are
nondynamical, which can be best understood from the ADM
formalism~\cite{adm}. In the ADM formalism, the gravitational
action is decomposed into the dynamical part containing the
spatial metric $h_{ij}$ and a part containing the lapse (N) and
shift functions ($N_i$). The lapse and shift functions have no
kinetic terms in the action, and they can be integrated out by
solving their equations of motion~\footnote{The equations of
motion derived from extremezing the gravitational action with
respect to lapse and shift functions result in the momentum and
hamiltonian constraints.}. This leaves only $h_{ij}$ as dynamical
modes, which have only two degrees of freedom. In summary, the
metric perturbations have only 2 dynamical degrees of freedom
(which are the gravitational wave polarizations in the standard
case) and 4 nondyncamical modes. For the case of the vector field,
out of 4 perturbations to start with ($\delta A_{\mu}$), one
perturbation can be removed by the $U(1)$ gauge transformation.
Out of the remaining 3 perturbations 2 of them are dynamical and
one mode is nondyamical ($\delta A_0$). The scalar field
perturbation introduces a single dynamical degree of freedom.
Therefore, in total, the perturbations comprise of 5 dynamical
modes, 2 of which are $2d$ vector and 3 are $2d$ scalar modes.
When additional scalar fields are considered, each field
introduces an extra scalar dynamical degree of freedom.

We will insert the perturbative expansion of the metric, the
vector field and the scalar field(s) into the starting action and
expand it at the quadratic order. We will show that the actions for
the $2d$ vector and $2d$ scalar modes are decoupled, so we study
them separately. We will determine the linear combinations of
perturbations that canonically normalize the action, generalizing
the computation of the standard isotropic case~\cite{musa}. The
study of the quadratic action is crucial for both showing that the
model is consistent (free of ghosts) and also for determining the
initial conditions for the perturbations. For modes that are
inside the horizon, canonical combinations of perturbations can be
quantized, and initial conditions can be set by the canonical
commutation relations and by the requirement that the adiabatic
vacuum state has minimal energy. We then proceed by numerically
integrating the evolution equations for the canonical fields, starting from
adiabatic initial conditions deep inside the horizon, until the
end of inflation, which gives us the primordial power spectra. We
do so for the $2d$ vector modes in this paper (which will be
related to the power spectrum of the $h_{\times}$ polarization of
gravitational waves) and perform the study of the spectrum of $2d$
scalar modes in a separate publication.

The paper is organized as follows: In section~\ref{sec:back}, we
summarize the anisotropic inflationary background solution
obtained in reference~\cite{Watanabe:2009ct} and generalize the
model to possess extra scalar fields. This generalization leads to
isotropization before the end of inflation. In
section~\ref{sec:perturbations}, we discuss the perturbations
around the background configuration. Specifically, in
subsection~\ref{sub:decomposition}, we review and discuss the
decomposition of perturbations around the Bianchi-I background
solution, following the formulation of reference~\cite{gcp2}. In
subsection~\ref{sub:generic}, we discuss the generic properties of
coupled vector-scalar models and develop tools that we will use in
the following sections in order to find adiabatic initial
conditions for such systems.

In subsection~\ref{sub:2dV} we discuss $2d$ vector perturbations
around the background configuration. We compute the action for
$2d$ vector modes and find the combinations of perturbations that
canonically normalize the action. In section~\ref{sec:power}, we
compute the power spectra of the $2d$ vector modes numerically.
This study results in the spectrum of $h_{\times}$ gravitational
wave mode. We show that the spectrum has angular dependence (so
breaks statistical isotropy) at large scales and reduces to the
standard nearly scale invariant form at small scales. We also
provide a fit to the numerically obtained spectrum. Finally in
section~\ref{sec:conclusions} we provide a general discussion
about the results we have obtained and their possible relation to
observations.

We also provide two extensive appendices: In
appendix~\ref{sub:app1}, we derive the relation between the $2d$
decomposed modes and the standard longitudinal mode, which has
been used to determine the power spectra. In
appendix~\ref{sub:app2}, we perform the study of the $2d$ scalar
modes. More precisely, we compute the quadratic action and
determine the modes that canonically normalizes the action. The spectrum of scalar perturbations
and the consequent phenomenological predictions (specifically, the $<a_{lm}\, a_{l'm'}^*>$ correlation)
will be presented elsewhere.

\section{Anisotropic inflationary expansion due to coupled vector and scalar fields}
\label{sec:back}

The models we discuss in this work are generically described by
the following action
\begin{equation}
S = \int\, d^4x\, \sqrt{-g}\, \left[ \frac{M_p^2}{2}\, R -
\sum_{a=1}^{N}\, \left( \frac{1}{2}\, \partial_{\mu} \phi_a\,
\partial^{\mu} \phi_a + V_a(\phi_a) \right) - \frac{1}{2}\,
\partial_{\mu} \phi\, \partial^{\mu} \phi - V(\phi) -
\frac{1}{4}\, f^2(\phi)\, F_{\mu\nu}\, F^{\mu\nu} \right]
\label{action}
\end{equation}
The original discussion of these coupled vector-scalar field
theories in an anisotropic inflationary background was given
in~\cite{Watanabe:2009ct}, where there is only the single scalar
field $\phi$ in the action. We assume that only one of the scalar fields
(or only one linear combination) enters in the kinetic function $f$ for the 
vector field. We denote the remaining scalars with $\phi_a$ $(a=1, \dots ,N)$.
The metric ansatz is taken to be the
homogeneous but anisotropic Bianchi-I metric given by
\begin{equation}
ds^2 = -dt^2 + e^{2 \alpha(t)}\, \left[ e^{-4\sigma(t)}\, dx^2 +
e^{2 \sigma(t)}\, \left( dy^2 + dz^2 \right) \right] \label{metric}
\end{equation}
In the above metric $\alpha$ measures the number of e-folds of
average isotropic expansion of the universe and $\sigma$ measures
the deviation from anisotropy. This metric also possesses a left
over $2d$ isotropy in the $y-z$ plane (which we will make use of
when decomposing the perturbations). We can also identify
(\ref{metric}) with $ds^2 = -dt^2 + a^2\, dx^2 + b^2\, (dy^2 +
dz^2)$, then we would have
\begin{equation}
a = e^{\alpha-2 \sigma} \,\,\, , \,\,\, b = e^{\alpha + \sigma}
\,\,\, , \,\,\, H \equiv \frac{H_a + 2 H_b}{3} = \dot{\alpha}
\,\,\, , \,\,\, h \equiv \frac{H_b - H_a}{3} = \dot{\sigma}
\end{equation}
where $H_a=\dot{a}/a$, $H_b=\dot{b}/b$, $H$ is the average
isotropic expansion rate, and $h$ is the anisotropic expansion
rate. An overdot denotes derivative with respect to the cosmic
time $t$. An isotropic flat FRW universe is the limiting case when
$h\rightarrow 0$. We will make use of both representations of the
Bianchi-I metric in the following discussions. The field equations
obtained from the action (\ref{action}) are
\begin{eqnarray}
&& G_{\mu\nu} = \frac{1}{M_p^2}\, \Bigg[ \partial_{\mu} \phi\,
\partial_{\nu} \phi - g_{\mu\nu}\, \left( \frac{1}{2}\,
\partial_{\alpha} \phi\, \partial^{\alpha} \phi +
V\left(\phi\right) \right) + \sum_{a=1}^{N} T_{\mu\nu}^{(a)}
\nonumber\\
&& \qquad\qquad\qquad + f^2\left(\phi\right)\, F_{\mu}^{\,\,
\alpha}\, F_{\nu\, \alpha} - \frac{1}{4}\, g_{\mu\nu}\,
f^2\left(\phi\right)\, F_{\alpha\,
\beta}\, F^{\alpha\, \beta} \Bigg] \nonumber\\
&& \frac{1}{\sqrt{-g}}\, \partial_{\mu}\, \left[ \sqrt{-g}\,
g^{\mu\nu}\, \partial_{\nu}\, \phi \right] - V'\left(\phi\right) -
\frac{f(\phi)\, f'(\phi)}{2}\, F_{\alpha\, \beta}\, F^{\alpha\,
\beta} = 0 \nonumber\\
&& \frac{1}{\sqrt{-g}}\, \partial_{\mu}\, \left[ \sqrt{-g}\,
g^{\mu\nu}\, \partial_{\nu}\, \phi_a \right] - V_a'\left(\phi_a\right) = 0 \nonumber\\
&& \frac{1}{\sqrt{-g}}\, \left[ \sqrt{-g}\, f^2\left(\phi\right)\,
F^{\mu\nu} \right] = 0 \label{fldeqns}
\end{eqnarray}
where
\begin{equation}
T_{\mu\nu}^{(a)} = \partial_{\mu} \phi_a\, \partial_{\nu} \phi_a -
g_{\mu\nu}\, \left( \frac{1}{2}\, \partial_{\alpha}\phi_a\,
\partial^{\alpha}\phi_a + V_a(\phi_a) \right)
\end{equation}
and $f'(\phi) \equiv df/d\phi$. We assume that all the scalar
fields are homogeneous so that $\phi=\phi(t)$ and
$\phi_a=\phi_a(t)$ for all $a$. The vector field is assumed to
have a homogeneous vacuum expectation value along the
$x$-direction so $A_{\mu} = \left( 0, \, A_1(t), \, 0, \, 0
\right)$ (a homogeneous $A_0$ does not enter into the field equations, and can
be set to zero by using the $U(1)$ gauge invariance). With the
metric and vector field ansatz, the last of (\ref{fldeqns}) can
be integrated to give
\begin{equation}
\dot{A}_1 = f^{-2}\left(\phi\right)\, p_A\, e^{-\alpha - 4 \sigma}
\label{solA}
\end{equation}
where $p_A$ is an integration constant. We use (\ref{solA}) and
the metric ansatz (\ref{metric}) in the field equations
(\ref{fldeqns}) to obtain the evolution equations
\begin{eqnarray}
&& 3 \dot{\alpha}^2 - 3 \dot{\sigma}^2 = \frac{1}{M_p^2}\, \left[
\frac{1}{2}\, \dot{\phi}^2 + V\left(\phi\right) + \sum_{a=1}^{N}\,
\rho_a(t) \right] + \frac{\tilde{p}_A^2}{2 M_p^2}\,
f^{-2}\left(\phi\right) \nonumber\\
&& 2 \ddot{\alpha} + 3 \dot{\alpha}^2 + 3 \dot{\sigma}^2 =
\frac{1}{M_p^2}\, \left[ - \frac{1}{2}\, \dot{\phi}^2 +
V\left(\phi\right) - \sum_{a=1}^{N}\, p_a(t) \right] -
\frac{\tilde{p}_A^2}{6 M_p^2}\, f^{-2}\left(\phi\right)\nonumber\\
&& \ddot{\sigma} + 3 \dot{\alpha}\, \dot{\sigma} =
\frac{\tilde{p}_A^2}{3 M_p^2}\, f^{-2}\left(\phi\right) \nonumber\\
&& \ddot{\phi} + 3 \dot{\alpha}\, \dot{\phi} + V'\left(\phi\right)
- \tilde{p}_A^2\, f^{-3}\left(\phi\right)\, f'\left(\phi\right) = 0 \nonumber\\
&& \ddot{\phi}_a + 3\, \dot\alpha\, \dot{\phi}_a + V_a'(\phi_a) =
0 \,\,\,\, , \,\,\,\, a=1,\ldots, N \label{evolution}
\end{eqnarray}
where we have defined
\begin{equation}
\rho_a \equiv \frac{1}{2}\, \dot\phi_a^2 + V_a(\phi_a) \,\,\,\, ,
\,\,\,\, p_a \equiv \frac{1}{2}\, \dot\phi_a^2 - V_a(\phi_a)
\,\,\,\, , \,\,\,\, \tilde{p}_A \equiv p_A\, e^{-2\,(\alpha +
\sigma)}
\end{equation}
The values of the functions $\alpha$ and $\sigma$ are nonphysical;
what is physical is their time derivatives. Therefore physics
is the same under a constant shift $\alpha(t) \rightarrow
\alpha(t) +\alpha_0 $ and $\sigma(t) \rightarrow \sigma(t) +
\sigma_0$ (This corresponds to a rescaling of the spatial
coordinates by two constant factors.). Clearly, the value of the
constant $p_A$ is nonphysical, and it changes under a coordinate
transformation. We can deduce its transformation properties by
imposing invariance of $F_{\alpha\beta}\, F^{\alpha\beta}$ which
reduces to the condition
\begin{equation}
F_{\alpha\beta}\, F^{\alpha\beta} \rightarrow F_{\alpha\beta}\,
F^{\alpha\beta}\,\,\,\,\, {\rm under}\,\,\,\, \{\alpha \rightarrow
\alpha + \alpha_0 , \,\, \sigma \rightarrow \sigma + \sigma_0\}
\,\,\, {\rm} \Rightarrow \,\,\,\, , \,\,\,\, p_A \rightarrow p_A\,
e^{2\, (\alpha_0 + \sigma_0)}
\end{equation}
Thus, $\tilde{p}_A$ is a physical parameter invariant under the
constant coordinate rescaling. For this reason, it is the
combination which enters in the field equations (\ref{evolution}).

We will discuss inflationary solutions of (\ref{evolution}), with
appropriate choices of $f(\phi)$ for both single field (only
$\phi$) and two field models (a=1) respectively.

\subsection{Single Scalar Field Inflationary Background}
\label{sub:1field}

In this subsection, we review and discuss the background evolution
of the original model~\cite{Watanabe:2009ct} with a single scalar
field $\phi$ coupled to the vector field. As described
in~\cite{Watanabe:2009ct}, we look for solutions that have a
slow-roll regime and small anisotropy. Thus, we neglect
$\dot\sigma, \, \dot\phi$ in the first and $\ddot\alpha$ in the
last of (\ref{evolution}) to get
\begin{equation}
3 \dot\alpha^2 \approx \frac{V(\phi)}{M_p^2} +
\frac{\tilde{p}_A^2}{2 M_p^2}\, f^{-2}(\phi) \,\,\,\,\, , \,\,\,\,
3\, \dot\alpha\, \dot\phi \approx -V'(\phi) - \tilde{p}_A^2\,
f^{-3}(\phi)\, f'(\phi) \label{evolution-apprx}
\end{equation}
Up to now, $f(\phi)$ has been arbitrary. We now specify it so to
obtain the desired anisotropic background solution. To do so, we
first assume that the effect of the vector field is completely
negligible, and the resulting equations can be solved
approximately in the standard slow-roll approximation, given by
\begin{equation}
\alpha \approx -\frac{1}{M_p^2}\, \int\,
\frac{V(\phi)}{V'(\phi)}\, d\phi
\end{equation}
The case $\dot\sigma=0$ corresponds to the isotropic FRW
background. A prolonged anisotropic inflation requires $\dot{A}_1$
to be nearly constant. From equation (\ref{solA}) we see that this
can be achieved if we choose
\begin{equation}
f(\phi) \sim e^{-2\alpha} \sim {\rm Exp}\left[ \frac{2}{M_p^2}\,
\int\, \frac{V(\phi)}{V'(\phi)}\, d\phi \right]
\end{equation}
For instance, when $V\propto \phi^n$, this becomes $f \sim
e^{\phi^2/n M_p^2}$. Furthermore, one can set $f \sim e^{c
\phi^2/n M_p^2}$ (where $c$ is a constant) and for $c>1$, the
anisotropy is expected to grow. From now on, we assume that the
functional form of $f(\phi)$ is given by
\begin{equation}
f(\phi) = {\rm Exp}\left[ \frac{2 c}{M_p^2}\, \int
\frac{V(\phi)}{V'(\phi)}\, d\phi \right] \label{fphi}
\end{equation}
We now return back to equations (\ref{evolution-apprx}) and look
for solutions that have small anisotropy, but the vector field
contribution is not totally negligible. We specify the scalar
field potential as $V(\phi)=m^2\, \phi^2/2$ and set $f=e^{c\,
\phi^2/2 M_p^2}$ from now on. The vector field modifies the
evolution of the scalar field, if  the terms $V'(\phi)$ and
$\tilde{p}_A^2\, f^{-3}(\phi)\, f'(\phi)$ are comparable. Namely,
for our choice of the potential and $f(\phi)$
\begin{equation}
m^2 \sim \frac{c\, \tilde{p}_A^2}{M_p^2}\, e^{-c\, \phi^2/M_p^2}
\label{aux.eq1}
\end{equation}
For such type of solutions, we can compare the energy densities of
the scalar and vector fields. The ratio, (when the scalar field is
still in the slow-roll regime and the anisotropy is small) is
given by
\begin{equation}
\frac{\rho_A}{\rho_\phi} \approx \frac{\tilde{p}_A^2\, e^{-c\,
\phi^2/M_p^2}}{m^2\, \phi^2} \approx \frac{M_p^2}{c\, \phi^2}
\end{equation}
where, the second approximate equality is obtained from
(\ref{aux.eq1}). This ratio is much smaller than $1$ during
inflation, since $\phi/M_p \gg 1$. Thus, even when the dynamics of
the scalar field is modified by the action of the vector field, as
long as the anisotropy is kept small enough, the energy density of
the vector field can be neglected. Therefore, we can rewrite
equations (\ref{evolution-apprx}) by neglecting the effect of the
vector field only in the first equation as
\begin{equation}
\dot\alpha^2 \approx \frac{m^2\, \phi^2}{6 M_p^2} \,\,\,\,\, ,
\,\,\,\, 3\, \dot\alpha\, \dot\phi \approx -m^2\, \phi + \frac{c\,
\tilde{p}_A^2}{M_p^2}\, \phi\, e^{-c\, \phi^2/M_p^2}
\label{evolution-apprx-2}
\end{equation}
These equations are integrated to give
\begin{equation}
e^{c\, \phi^2/M_p^2 + 4\, \alpha} = \frac{c^2\, p_A^2}{m^2\,
M_p^2\, \left( c-1 \right)} + D\, e^{-4\, (c-1)\, \alpha}
\label{aux.eq2}
\end{equation}
where $D$ is an integration constant. When the second term in the
right hand side of the above equality is dominant we get
\begin{equation}
\frac{c\, \phi^2}{M_p^2} + 4\, c\, \alpha \sim {\rm constant}
\,\,\,\, \rightarrow \,\,\,\, \dot\phi \approx
-\sqrt{\frac{2}{3}}\, m\, M_p \label{region1}
\end{equation}
which is the standard slow-roll relation. The second term in the
right hand side of (\ref{aux.eq2}) will eventually be subdominant
for $c>1$, since the universe expands and $\alpha$
grows~\footnote{For $c \le 1$, the initial anisotropy will decay
exponentially fast so we do not consider this possibility here.
The solutions with $c \le 1$ are disconnected from the anisotropic
attractor solutions $c > 1$, that is there is no smooth limit of
$c\rightarrow 1$ which takes us back to the isotropic FRW
solution. We will discuss the implications of this effect on the
perturbations in the next section.}. Therefore, the first term
eventually dominates and we have
\begin{equation}
e^{c\, \phi^2/M_p^2} \approx \frac{c^2\, \tilde{p}_A^2}{m^2\,
M_p^2\, (c-1)} \,\,\,\, \rightarrow \,\,\,\, \dot\phi \approx
-\sqrt{\frac{2}{3}}\, \frac{m\, M_p}{c} \label{region2}
\end{equation}
which has an extra $1/c$ factor compared to the standard slow-roll
solution (\ref{region1}). Using the third of (\ref{evolution}),
for small and slowly varying anisotropy we have
\begin{equation}
3\, \dot\alpha\, \dot\sigma \approx \frac{\tilde{p}_A^2}{3
M_p^2}\, e^{-c\, \phi^2/M_p^2} \label{region2-sigma}
\end{equation}
Moreover, combining this result with (\ref{region2}) and the first
of (\ref{evolution-apprx-2}) we find that the anisotropy obeys
\begin{equation}
\frac{\dot\sigma}{\dot\alpha} = \frac{\tilde{p}_A^2\, e^{-c\,
\phi^2/M_p^2}}{9\, M_p^2\, \dot\alpha^2} \approx \frac{2}{3}\,
\frac{M_p^2\, (c-1)}{c^2\, \phi^2} \propto
\frac{\rho_A}{\rho_{\phi}} \ll 1 \label{sigmadot}
\end{equation}
which is always satisfied (either in the region described by
(\ref{region1}) or (\ref{region2})). Also, note that the
anisotropy is slowly increasing with time, since $\phi$ is
decreasing towards the end of inflation. The slow-roll solutions
described by (\ref{region1}) is the standard isotropic
inflationary attractor and the solutions described by
(\ref{region2}) is the anisotropic inflationary attractor. We are
interested in configurations that start close to the attractor
solution (\ref{region2}). For illustration, we solve the system
(\ref{evolution}) numerically by setting the initial conditions on
the attractor solution. Namely, we set initial conditions as
\begin{eqnarray}
&& \dot\phi_{in} = -\frac{m^2}{3 c\, \dot\alpha_{in}}\, \phi_{in}
\nonumber\\
&& \dot\sigma_{in} = \frac{m^2\, (c-1)}{9 c^2\, \dot\alpha_{in}} \nonumber\\
&& \frac{\tilde{p}_{A\, in}^2}{M_p^2\, f^2(\phi)} = \frac{m^2\,
(c-1)}{c^2} \label{initcond}
\end{eqnarray}
Inserting (\ref{initcond}) into the constraint equation (the first
of (\ref{evolution}), we solve for $\dot\alpha_{in}$ to get
\begin{equation}
\dot\alpha_{in}^2 = \frac{m^2}{12}\, \left[ \frac{c-1}{c^2} +
\frac{\phi_{in}^2}{M_p^2} + \sqrt{\frac{25 (c-1)^2}{9 c^4} +
\frac{2 (1+3 c)}{3 c^2}\, \frac{\phi_{in}^2}{M_p^2} +
\frac{\phi_{in}^4}{M_p^4}}\, \right] \label{ain}
\end{equation}
Equations (\ref{initcond}) and (\ref{ain}) specifies the slow-roll
initial conditions, which are set on the attractor solution. The
only free parameters are $\phi_{in}, \, \alpha_{in}$ and
$\sigma_{in}$. The values of $\alpha_{in}$ and $\sigma_{in}$ are
nonphysical, so we can set $\alpha_{in}=\sigma_{in}=0$ for
convenience (which is equivalent to setting $a_{in}=b_{in}=1$).
Thus, only $\phi_{in}$ needs to be specified, which determines the
amount of inflation. The number of e-folds of inflation is
approximately given by $\alpha(t_{end}) = N \approx c\,
\phi_{in}^2/4 M_p^2$, so for $c=2$ we set $\phi_{in}=11 M_p$ which
gives around $60$ e-folds of inflation.
\begin{figure}[h]
\centerline{
\includegraphics[width=0.5\textwidth]{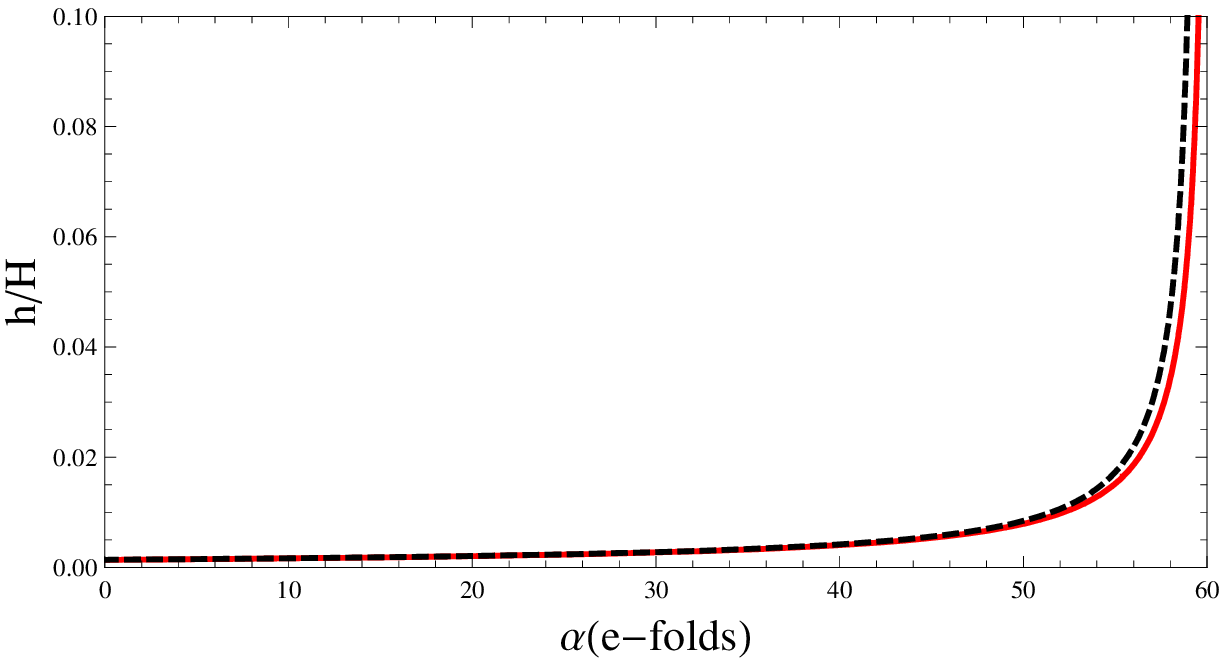}
\includegraphics[width=0.5\textwidth]{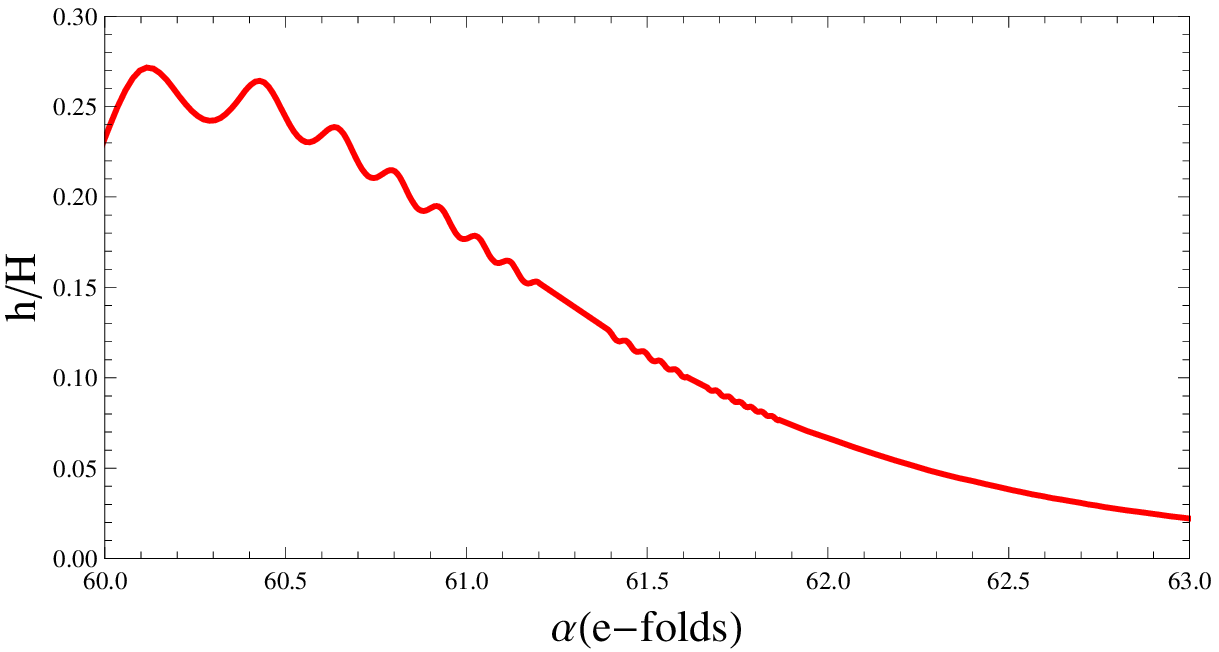}
} \caption{The left panel shows the evolution of anisotropy during
inflation and the right panel after inflation (during when the
scalar field is oscillating). $H \equiv \dot\alpha$ and $h \equiv
\dot\sigma$. The black dashed curve represents the approximate
slow-roll solution during inflation and the straight red curves the numerical
solution.} \label{fig:plot1}
\end{figure}
\begin{figure}[h]
\centerline{
\includegraphics[width=0.5\textwidth]{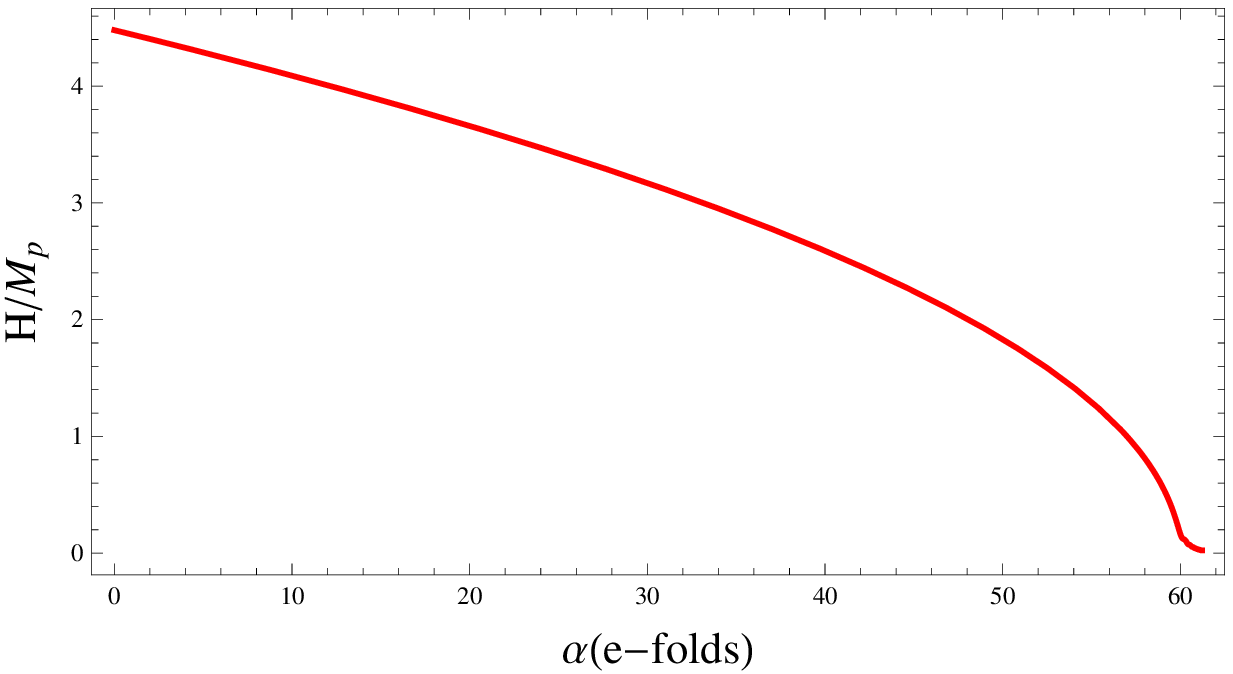}
\includegraphics[width=0.5\textwidth]{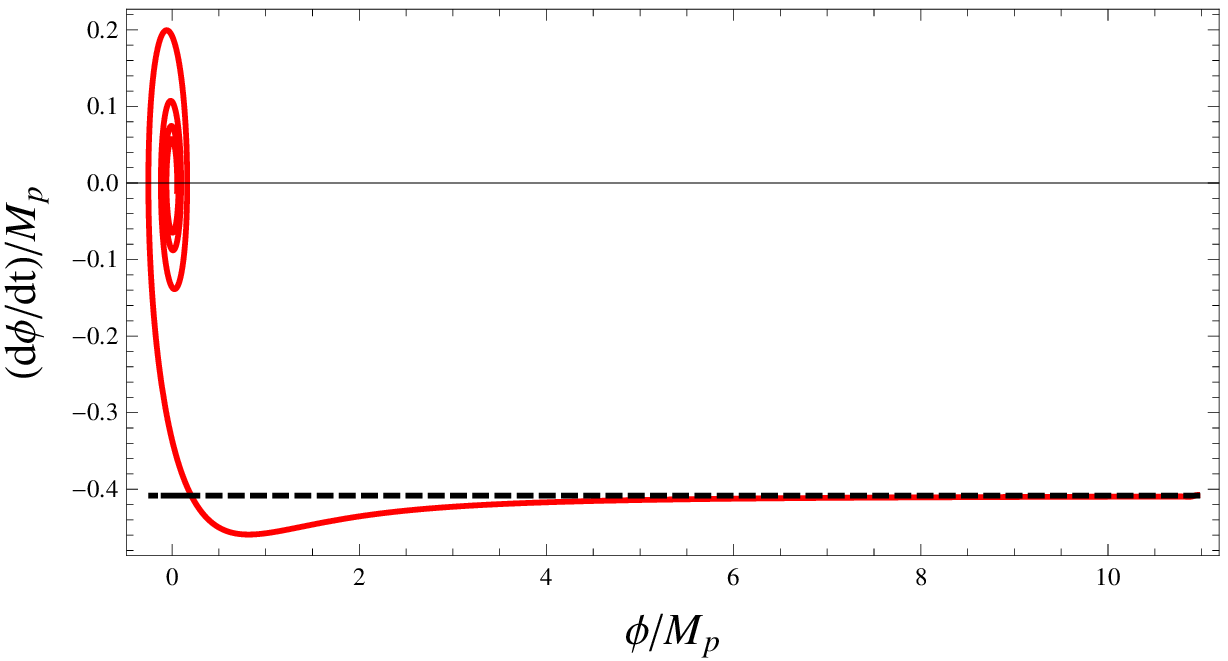}
} \caption{The left panel shows the evolution of the average
Hubble rate ($H\equiv \dot\alpha$) and the right panel is the
phase plot of the scalar field. Again, the black dashed lines are
the slow-roll solutions. Time is measured in units of $m$.} \label{fig:plot2}
\end{figure}
We show in Figs.~\ref{fig:plot1},~\ref{fig:plot2} the numerical
evolution of the system. Note that the slow-roll and the numerical
solutions are in good agreement with each other. The anisotropy in
the single field model is increasing towards the end of inflation
and decreases after inflation (as can be seen in the right panel
of Fig.~\ref{fig:plot1}) during when the scalar field has an
effective equation of state of matter. We are however interested
in cases where the anisotropic expansion takes place at the onset
of inflation (for example as in~\cite{gcp,gcp2,uzan1,uzan2}
and~\cite{soda1}). The main reason is because we only want the
largest observable scales to be modified from the perturbations
that left the horizon during the anisotropic initial stage. This
in turn might have some relevance to the alignment of the lowest
multipoles. In the current case, all observable scales are
modified (indeed smaller scales are modified even more since the
anisotropy increases towards the end of inflation) and this is not
compatible with observations. A simple possibility to overcome
this difficulty is to introduce extra scalar fields as in the
starting action (\ref{action}). We discuss the background
evolution for the simplest possibility of a two-field modification
of the original proposal of~\cite{Watanabe:2009ct} in the next
subsection.

\subsection{Two Scalar Field Inflationary Background}
\label{sub:2field}

In this section we discuss the simplest multi-field case of the
action (\ref{action}) with $a=1$. In this model, the extra scalar
field $\phi_1$ is not coupled to the vector field, and it is only
relevant for the overall isotropic expansion of the universe. For
the rest of the paper we will assume that $V(\phi) = m^2\,
\phi^2/2$ and $V_1(\phi_1)=m_1^2\, \phi_1^2/2$. The ratio of the
masses $\mu \equiv m_1/m$ is chosen to be smaller than 1. This
choice leads to a two stage inflationary expansion. The first
stage is driven mainly by $\phi$, and due to its coupling with the
vector field, this stage is anisotropic and it proceeds in a
similar fashion as in the previous subsection. At the end of the
first stage, the field $\phi_1$ takes over the expansion, while
the field $\phi$ starts oscillating. Since the expansion of the
universe still proceeds, the Hubble friction will drive the field
$\phi$ to zero and consequently, the coupling $f(\phi)$ approaches
unity. Therefore, from (\ref{solA}), we clearly see that the
effect of the vector field quickly diminishes. Thus, after the
first stage, the universe quickly isotropizes and inflation
proceeds isotropically until the field $\phi_1$ starts
oscillating. Problems related to an inflationary stage
that has increasing anisotropy will be eliminated by the two stage modification.

\begin{figure}[h]
\centerline{
\includegraphics[width=0.5\textwidth]{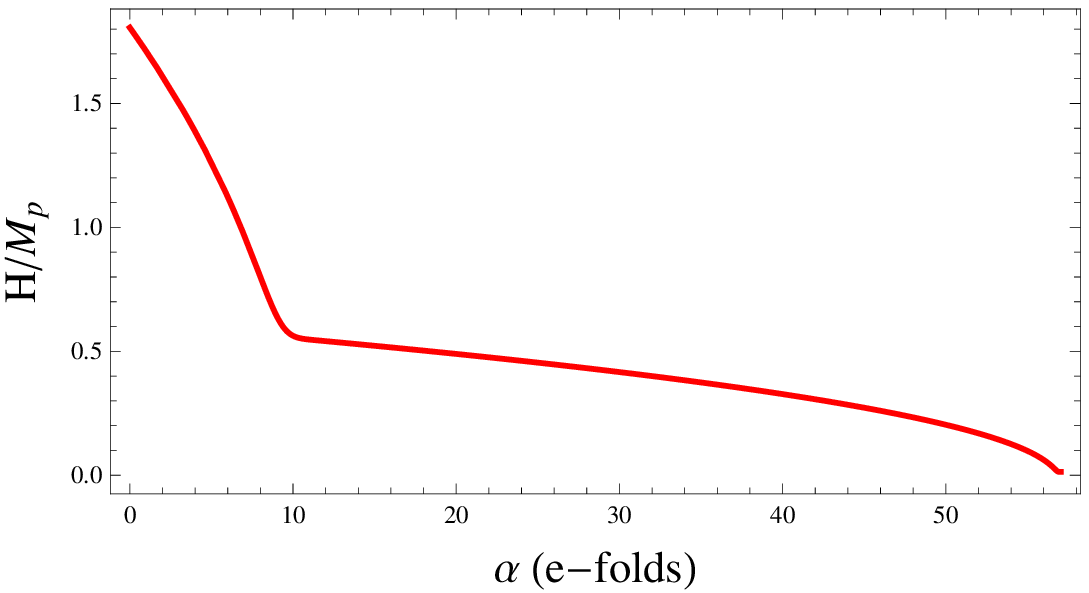}
\includegraphics[width=0.5\textwidth]{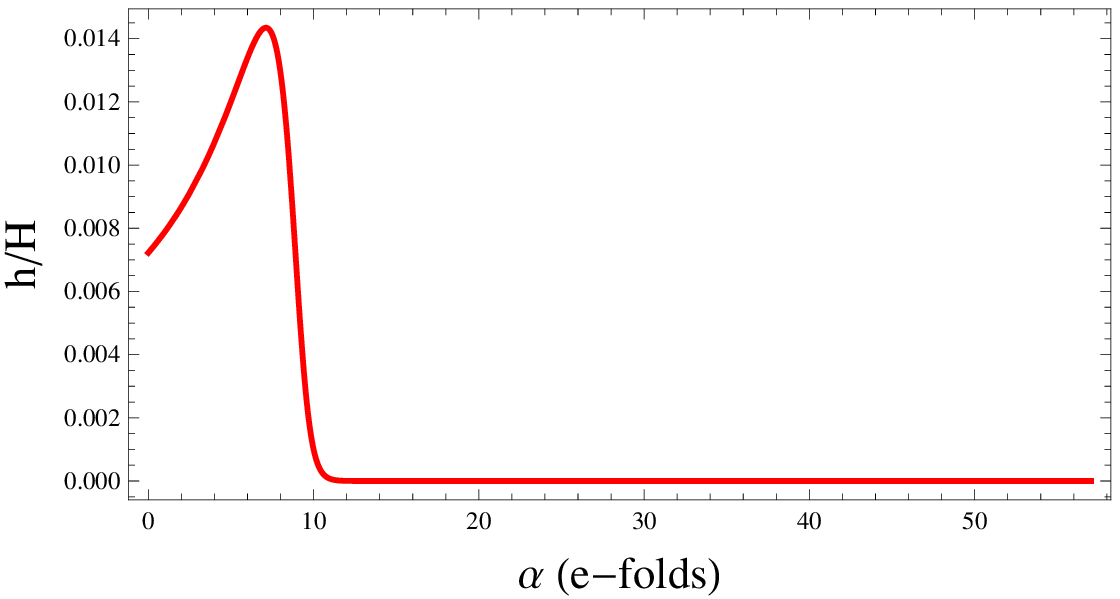}
} \caption{The left panel shows the evolution of the average
Hubble expansion. The right panel shows the evolution of
anisotropy, where $H \equiv \dot\alpha$ and $h \equiv
\dot\sigma$.} \label{fig:plot3}
\end{figure}

\begin{figure}[h]
\centerline{
\includegraphics[width=0.5\textwidth]{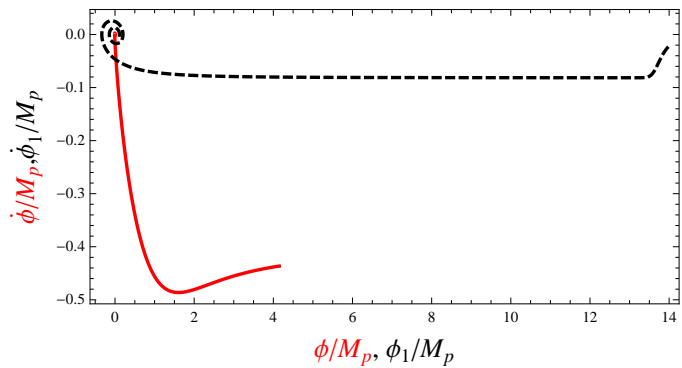}
} \caption{The phase portrait for the two scalar fields. The
straight red line represents $\phi$ and the dashed black line
represents $\phi_1$. Time is measured in units of $m$.} \label{fig:plot4}
\end{figure}

As we have done in the previous section, we first study the
solutions to the background evolution equations, which are given
in this case as
\begin{eqnarray}
&& 3 \dot\alpha^2 - 3 \dot\sigma^2 = \frac{1}{M_p^2}\, \left[
\frac{1}{2}\, \dot\phi^2 + \frac{1}{2}\, \dot\phi_1^2 +
\frac{1}{2}\, m^2\, \phi^2 + \frac{1}{2}\, \mu^2\, m^2\, \phi_1^2
\right] + \frac{\tilde{p}_A^2}{2 M_p^2\,
f^2(\phi)} \nonumber\\
&& 2 \ddot\alpha + 3 \dot\alpha^2 + 3 \dot\sigma^2 =
\frac{1}{M_p^2}\, \left[ - \frac{1}{2}\, \dot\phi^2 -
\frac{1}{2}\, \dot\phi_1^2 + \frac{1}{2}\, m^2\, \phi^2 +
\frac{1}{2}\, \mu^2\, m^2\, \phi_1^2 \right]
-\frac{\tilde{p}_A^2}{6 M_p^2\,
f^2(\phi)} \nonumber\\
&& \ddot\sigma + 3 \dot\alpha\, \dot\sigma =
\frac{\tilde{p}_A^2}{3
M_p^2\, f^2(\phi)} \nonumber\\
&& \ddot\phi + 3 \dot\alpha\, \dot\phi + m^2\, \phi -
\frac{\tilde{p}_A^2}{f^3(\phi)}\, f'(\phi) = 0 \nonumber\\
&& \ddot\phi_1 + 3 \dot\alpha\, \dot\phi_1 + m^2\, \mu^2\,
\phi_1^2 = 0 \label{evolution-2}
\end{eqnarray}
where $f(\phi)=e^{c\, \phi^2/2 M_p^2}$ as before. We can obtain
slow-roll solutions similar to (\ref{region2}), in the first stage
which proceeds anisotropically. Equations
(\ref{evolution-apprx-2}) are now modified to become
\begin{equation}
\dot\alpha^2 \approx \frac{m^2}{6 M_p^2}\, \left( \phi^2 +
\mu^2\, \phi_1^2 \right) \,\,\,\, , \,\,\,\, 3\, \dot\alpha\,
\dot\phi \approx -m^2\, \phi + \frac{c\, \tilde{p}_A^2}{M_p^2}\, \phi\, 
e^{-c\, \phi^2/M_p^2} \,\,\,\, , \,\,\,\, 3\, \dot\alpha\,
\dot\phi_1 \approx -m^2\, \mu^2\, \phi_1 \label{evolution-apprx-3}
\end{equation}
Initially, the field $\phi$ drives the expansion, so $\phi^2 \gg
\mu^2\, \phi_1^2$, and the slow roll solutions for the first stage
is simply given similar to (\ref{region2}) as
\begin{eqnarray}
&& e^{c\, \phi^2/M_p^2} \approx \frac{c^2\, \tilde{p}_A^2}{m^2\,
M_p^2\, (c-1)}\,\,\, \rightarrow \,\,\, \dot\phi \approx
-\sqrt{\frac{2}{3}}\, \frac{m\, M_p}{c} \nonumber\\
&& \dot\phi_1 \approx - \sqrt{\frac{2}{3}}\, \frac{m\, M_p\,
\mu^2}{\phi}\, \phi_1 \label{2f.region1}
\end{eqnarray}
and the  anisotropy is given precisely as in equation
(\ref{sigmadot}). After this first stage ends, the isotropic
expansion of the universe is driven by $\phi_1$ and this second
stage is described by the following slow-roll solution
\begin{equation}
\dot\phi_1 \approx - \sqrt{\frac{2}{3}}\, m\, M_p\, \mu
\label{2f.region2}
\end{equation}
We also integrate the system (\ref{evolution-2}) numerically and
the results are shown in Figs~\ref{fig:plot3}-~\ref{fig:plot4}. As
we have done in the previous section, we use the slow-roll
solutions and the constraint equation (the first of
(\ref{evolution-2})) to determine the initial conditions
$\dot\phi_{in}, \, \dot\phi_{1 in}, \, \dot\sigma_{in}, \,
\dot\alpha_{in}$~\footnote{In order to find the approximate
slow-roll solutions, one has to ignore the effect of the second
field $\phi_1$ initially. Although this approximation is fairly
accurate, we let the numerical evolution to run for a few e-folds
until the real attractor solution is reached.}. The initial
conditions for the scale factors are set as $\alpha_{in}=0$ and
$\sigma_{in}$ chosen to satisfy $\sigma(t_{end})=0$ at the end of
inflation. We have chosen $\phi_{in}=5 M_p$, $\phi_{1 in}=14 M_p$
and $\mu=1/10$, which gives around $60$ e-folds of inflation, with
the first $10$ e-folds being anisotropic.
In this type of background, when the anisotropic expansion takes
place at the onset of inflation, we expect that only the largest
scale perturbations are improved and at small scales standard
results are recovered.

\section{Perturbations}
\label{sec:perturbations}

This section is devoted to the study of the perturbations around
the anisotropic inflationary background solutions we have
discussed in the previous section. We will first discuss the
decomposition of perturbations around the background
configuration, which is different from standard
scalar-vector-tensor decomposition used for isotropic backgrounds.
The isotropic background solutions have a 3 dimensional rotational
and translational symmetry for constant time slices, therefore
small fluctuations are decomposed using the representations of the
$SO(3)$ group. In such a decomposition, one identifies decoupled
classes of fluctuations and according to their transformation
properties, one can identify scalar, vector (transverse) and
tensor modes(transverse-traceless). Such a decomposition can still
be considered in an anisotropic background, however fluctuations
classified as scalar, vector and tensor would be coupled in the
linearized calculation~\footnote{We refer to both the computation
of the quadratic action and linearized Einstein equations as the
"linearized calculation", since the latter follows from the
previous.} (see for example~\cite{uzan1,uzan2}). Instead, we make
use of the left over 2 dimensional rotational symmetry of the
$y-z$ plane of the Bianchi-I metric (\ref{metric}) in order to
decompose small fluctuations around the background as
in~\cite{gcp2,hcp2}. In such a two dimensional decomposition, there are
two decoupled sets of perturbations, the $2d$ scalar and $2d$
vector modes. We will compute the quadratic action of the small
fluctuations for both type of modes and study them separately.
Since the universe isotropizes either after inflation in the
single field model, or after the first stage of anisotropy in the
double field model, the observables today are best defined if we
use standard isotropic definitions of fluctuations. Therefore, we
will also find the map between the $2d$ decomposition and the
standard $3d$ isotropic decomposition, using which we will
determine the power spectra for the gravitational wave
polarizations $\{ h_{\times}, \, h_{+} \}$ and the curvature
perturbation ${\cal R}$ today (which are defined with respect to
an isotropic background). This procedure is described in detail in
the appendix.

\subsection{Decomposition of Perturbations}
\label{sub:decomposition}

We decompose the perturbations of the metric and the vector field
by exploiting the $2d$ isotropy of the background as
\begin{eqnarray}
\delta g_{\mu \nu} &=&
 \left( \begin{array}{ccc}  -2 \Phi & a\,
\partial_1\, \chi & b\left( \partial_i\, B + B_i \right) \\
 &  - 2 \, a^2 \, \Psi  & a\, b\, \partial_1\, \left(
 \partial_i\, \tilde{B} + \tilde{B}_i \right) \\
  & & b^2\, \left(  - 2\Sigma \, \delta_{ij} - 2\,
  \partial_i\, \partial_j\, E - \partial_i\, E_j - \partial_j\,
  E_i \right) \end{array} \right) \nonumber\\ \nonumber\\
 \delta A_\mu &=& \left( \alpha_0 ,\, \alpha_1 ,\, \partial_i \alpha + \alpha_i \right)
\label{defn-perts-2d}
\end{eqnarray}
Additionally, the system of perturbations contain the fluctuations
of the scalar field(s) $\delta\phi_{a}$. Indices $i,j=2,3$ span
the isotropic $2d$-plane. The perturbations $\{ \delta\phi_{a}, \,
\Phi, \, \chi, \, B, \, \Psi, \, \tilde{B}, \, \Sigma, \, E, \,
\alpha_0, \, \alpha_1, \, \alpha \}$ are $2d$ scalar
perturbations, comprising one degree of freedom (d.o.f) each, and
$\{ B_i, \, \tilde{B}_i, \, E_i, \, \alpha_i \}$ are $2d$ vector
modes which satisfy the transversality condition $(\partial_i\,
B_i=\dots=0)$. Therefore, the $2d$ vector perturbations also
comprise one d.o.f. each. The $2d$ vector and scalar modes are
decoupled at the linearized level.

We will work in the momentum space where a generic Fourier
transformed mode is defined as
\begin{equation}
\delta\left( x \right) = \frac{1}{\left( 2 \pi \right)^3}\, \int\,
d^3k\, \delta\left( k \right)\, e^{-i\, k_L\, x - i\, k_{T2}\, y -
i\, k_{T3}\, z} \label{fourier}
\end{equation}
We can use the residual $2d$-isotropy to fix the comoving momentum
to be aligned in the $x-y$ plane without any loss of generality.
Namely, we set $k_{T3}=0$ and $k_T=k_{T2}$ in what follows.
Therefore, in Fourier space, the generic $2d$ decomposition can be
written as (before fixing the gauge)
\begin{eqnarray}
\delta g_{\mu\nu}\, (k) &=& \left( \begin{array}{cccc} -2\, \Psi &
-i\, a\, k_L\, \chi & -i\, b\, k_{T}\, B & b\, B_3 \\
 & -2\, a^2\, \Psi & -a\, b\, k_L\, k_T\, \tilde{B} & -i\, a\, b\,
 k_L\, \tilde{B}_3 \\
  & & b^2\, \left( -2\, \Sigma + 2 k_T^2\, E \right) & i\, b^2\,
  k_T\, E_3 \\
   & & & -2\, b^2\, \Sigma \end{array} \right) \nonumber\\
\delta A_{\mu}(k) &=& \left( \alpha_0, \, \alpha_1, \, -i\, k_T\,
\alpha, \, \alpha_3 \right) \label{pert-k}
\end{eqnarray}
Notice that, for the $2d$ vector modes, fixing $k_T=k_{T2}$
resulted in $B_2 = \tilde{B}_2 = E_2 = \alpha_2 = 0$. Under the
infinitesimal coordinate transformation $x^{\mu} \rightarrow
x^{\mu} + \epsilon^{\mu}$, the metric perturbations transform as
\begin{equation}
\delta g_{\mu\nu} \rightarrow \delta g_{\mu\nu} - g_{\mu\nu,
\sigma}^{(0)}\, \epsilon^{\sigma} - g_{\mu\sigma}^{(0)}\,
\epsilon^{\sigma}_{,\nu} - g_{\sigma\nu}^{(0)}\,
\epsilon^{\sigma}_{,\mu}
\end{equation}
where we have decomposed
\begin{equation}
\epsilon^{\mu}(k) = \left( \xi^0, \, -i\, k_L\, \xi^1, \, -i\,
k_T\, \xi, \, \xi^3 \right)
\end{equation}
We choose to fix the coordinate gauge freedom by setting
$\tilde{B}=\Sigma=E=E_3=0$, namely we solve for the infinitesimal
transformation to satisfy
\begin{eqnarray}
&& \xi^0 = -\frac{1}{H_b}\, \Sigma \,\,\,\, , \,\,\,\, \xi = -E
\,\,\,\, , \,\,\,\, \xi^1 = \frac{b}{a}\, \left( \tilde{B} +
\frac{b}{a}\, E \right) \,\,\,\, , \,\,\,\,\xi^3 = -E_3
\label{ourgauge}
\end{eqnarray}
Since the action is invariant under the $U(1)$ gauge
transformation, we can exploit this symmetry to eliminate one of
the vector field perturbations. Namely, using the invariance
$\delta A_{\mu} \rightarrow \delta A_{\mu} + \partial_{\mu}\,
\lambda$, we set $\alpha=0$ in (\ref{defn-perts-2d}). This choice
completely fixes the $U(1)$ gauge freedom. In summary, the
complete system of perturbations, after fixing the coordinate and
$U(1)$ gauge freedoms, contain three $2d$ vector modes $\{
\tilde{B}_3, \, \alpha_3, \, B_3 \}$ and six $2d$ scalar modes $\{
\Phi, \, \chi, \, B, \, \Psi, \, \alpha_0, \, \alpha_1 \}$. We
will also have additional scalar modes coming from the
fluctuations of the scalar field(s) $\delta\phi_{a}$. We will show
in the following section that, not all the perturbations are
dynamical, namely some perturbations enter into the quadratic
action without any time derivatives. Such modes are lagrange
multipliers and they can be integrated out from the
action~\cite{adm}. The modes $\{ \tilde{B}_3, \, \alpha_3, \,
\Psi, \, \alpha_1 \}$ and perturbations of the scalar fields are
dynamical modes. On the other hand, $\{B_3, \, \Phi, \, \chi, \,
B, \, \alpha_0 \}$ are nondynamical modes as we will demonstrate.

As we have mentioned earlier, it is useful to find the connection
between the perturbations defined with respect to the $2d$
decomposition and perturbations defined with respect to standard
isotropic decompositions. This is useful in discussing the phenomenological
consequences. After the universe isotropizes, perturbations evolve 
in the standard way, what is different is that their solutions are 
nonstandard. However, it is useful to give results in terms of the 
modes which are commonly used in FRW studies. For definiteness, we find the
transformation between our choice of the gauge given in equation
(\ref{ourgauge}) and the standard longitudinal gauge. Here we
present the results and leave the details to the appendix. The
perturbations in the $2d$ decomposition are related to the
perturbations of the longitudinal gauge as
\begin{eqnarray}
\tilde{B}_3 &=& \frac{i}{k\, k_L}\, \left( \frac{k_T^2 + b/a\,
k_L^2}{k_T} \right)\, h_{\times} \nonumber\\
\Psi &=& \left( 1 - \frac{H_a}{H_b} \right)\, \psi - \left\{
\frac{H_a}{2 H_b} + \frac{1}{k^2}\, \left[ \frac{1}{2}\, k_T^2 +
\frac{b^3}{a^3}\, k_L^2\, \left( 1 + \frac{b}{a}\, \frac{2
k_L^2+k_T^2}{2 k_T^2} \right)\, \right]\, \right\}\, h_{+}
\nonumber\\
\delta\phi &=& \delta\phi^L + \frac{\dot\phi}{H_b}\, \left( \psi +
\frac{1}{2}\, h_{+} \right). \label{long-2d-metric}
\end{eqnarray}
In the above equations, the terms in the left hand side are
perturbations defined with respect to the $2d$ decomposition and
the terms in the right hand side are perturbations in the
longitudinal gauge, where $h_{\times}, \, h_{+}$ are the two
gravitational wave polarizations, $\psi$ is the spatial scalar
perturbation and $\delta\phi^{L}$ is the scalar field perturbation
(any generic scalar field in the model). These relations are
useful in computing the power spectra of the gravitational waves
and curvature perturbation.

\subsection{General Properties of Coupled System of Perturbations}
\label{sub:generic}

In this subsection, we discuss the general properties of the
coupled system of perturbations that arise in this study. Our main
discussion is based on the quadratic action for perturbations,
which we need in order to determine initial conditions by
canonical quantization. We compute the action for perturbations in
quadratic order by inserting the decompositions
(\ref{defn-perts-2d}), $\phi = \phi_0 + \delta\phi$ and $\phi_a =
\phi_{a0} + \delta\phi_a$ into (\ref{action}), after fixing the
$U(1)$ and coordinate gauges as described in the previous
subsection. We note that the $2d$ scalar and $2d$ vector
perturbations decouple in the quadratic action. We will provide
explicit expressions for the scalar and vector actions in the
following sections. Next we integrate out the nondynamical modes
from the action, to obtain an action in terms of the dynamical
modes only. More specifically, we extremize the quadratic action
with respect to the nondynamical modes and obtain equations of
motions for them (which are actually constraint equations). Then
we solve for the nondynamical modes from their equations of
motion, and insert the solutions back into the starting action.
The final action we obtain depends only on the dynamical modes.
The next step we perform is to find linear combinations of modes
that canonically normalize the quadratic action. More
specifically, the quadratic action after integrating out the
nondynamical modes is formally of the type $S \sim \int dt\,
d^3k\, ( \dot\delta_i \, K_{ij}\, \dot\delta_j + \dot\delta_i\,
\Lambda_{ij}\, \delta_j - \delta_i\, \Omega^2_{ij}\, \delta_j )$,
where $\delta_i$ denotes the dynamical modes. We find linear
combinations $\delta_i=R_{ij}\, \delta_{cj}$ such that
$\dot\delta_i\, K_{ij}\, \dot\delta_{j} = \frac{1}{2}\,
\dot\delta_{ci}\, \dot\delta_{ci} + \ldots$ where the dots denote
mixed terms. The modes $\delta_{ci}$ are the canonical modes. At
the moment we will study a generic action obtained after
canonically normalizing the starting action, which will be
relevant for both $2d$ vector and scalar modes. This action reads
\begin{equation}
S_{can} = \frac{1}{2}\, \int\, dt\, d^3k\, \left\{
\dot{\bf\Phi}^{\dag}\, \dot{\bf\Phi} + \left( i\,
\dot{\bf\Phi}^{\dag}\, {\bf X}\, {\bf\Phi} + {\rm h.c.} \right) -
p^2\, {\bf\Phi}^{\dag}\, \left( {\bf 1} + \frac{\omega^2}{p^2}
\right)\, {\bf\Phi} \right\} \label{genericaction}
\end{equation}
where ${\bf\Phi}$ is a column vector made from the canonically
normalized perturbations, ${\bf X}$ is a symmetric matrix and
${\bf\omega}^2$ is a hermitian matrix. In the above action, $p$ is
the physical momentum with $p^2=p_L^2+p_T^2$ where $p_L=k_L/a$ and
$p_T=k_T/b$. We would like to perform a unitary transformation in
the field space in order to get rid of the mixed terms ${\dot{\bf
\Phi}}^{\dag}\, X\, {\bf\Phi}$ and reduce the action to a canonical
form that can be used to quantize the fields (and in turn obtain
the initial conditions). This can be achieved by a unitary matrix
${\bf U}$ so that ${\bf\Phi}={\bf U}\, {\bf\Psi}$ (The mixed terms
can also be eliminated in the level of the Hamiltonian by
canonical transformations. See for example~\cite{Grisa:2009yy}).
We choose the matrix ${\bf U}$ to satisfy
\begin{equation}
{\dot{\bf U}} + i\, {\bf X}\, {\bf U} = 0 \label{eqnU}.
\end{equation}
Inserting the transformation back into (\ref{genericaction}) and
using (\ref{eqnU}) (with the unitarity condition on ${\bf U}$), we
get
\begin{equation}
S = \frac{1}{2}\, \int\, dt\, d^3k\, \left[ \dot{\bf
\Psi}^{\dag}\, \dot{\bf \Psi} - {\bf \Psi}^{\dag}\, {\bf
U}^{\dag}\, \left( p^2 + \omega^2 + {\bf X}^{\dag}\, {\bf X}
\right)\, {\bf U}\, {\bf \Psi} \right]. \label{actionpsi}
\end{equation}
We will show in the following sections that the matrix ${\bf X}$
is of the order of the average Hubble scale $H$ (i.e. ${\bf X} =
{\rm O}(H)$) and the matrix $\omega^2$ is of the order of Hubble
scale squared (i.e. $\omega^2 = {\rm O}(H^2)$). This means that in
the deep UV/subhorizon regime, where $p \gg H$, these matrices
will become negligible compared to $p^2$. Therefore, for modes
that are deep inside the horizon initially, the above action
reduces to a form which has diagonal mass terms
\begin{equation}
S = \frac{1}{2}\, \int\, dt\, d^3k\, \left[ \dot{\bf
\Psi}^{\dag}\, \dot{\bf \Psi} - p^2\, {\bf \Psi}^{\dag}\, {\bf
\Psi} + {\rm O}(H^2) \right]
\end{equation}
In this limit when the mass terms are diagonal, the mode expansion
for field ${\bf\Psi}$ simply reads
\begin{equation}
\hat{\bf \Psi}(x) = \int\, \frac{d^3k}{\left( 2\pi\right)^{3/2}}\,
\left[ e^{i\, \overrightarrow{k} \cdot \overrightarrow{x}}\,
v(t)\, \hat{a}(k) + e^{-i\, \overrightarrow{k} \cdot
\overrightarrow{x}}\, v^*(t)\, \hat{a}^{\dag}(k) \right]
\label{psiquant}
\end{equation}
The creation and annihilation operators $\hat{a}(k),\,
\hat{a}^{\dag}(k)$ obey the standard bosonic commutation relations
and the mode functions $v(t)$ satisfy
\begin{equation}
\ddot{v} + p^2\, v = 0. \label{eomv}
\end{equation}
The above equation can be solved using the WKB approximation in
the adiabatic limit, which gives
\begin{equation}
v(t) = \frac{1}{\sqrt{2 \omega}}\, e^{-i\, \int\, \omega\, dt'}
\end{equation}
where $\omega$ is given by
\begin{eqnarray}
\omega &=& p - \frac{\ddot{p}}{4 p^2} + \frac{3}{8}\,
\frac{\dot{p}^2}{p^3} + \dots \nonumber\\
&=& p + O(H^2)
\end{eqnarray}
In the above equation, $\dots$ represent higher order corrections
which are suppressed by higher powers of $H$ in the adiabatic
solution. This is also consistent with the accuracy of the mode
expansion (\ref{psiquant}). It is of course possible to neglect 
the mixed terms proportional to ${\bf X}$ and
$\omega^2$ from the beginning. In this way the action for the original
fields ${\bf \Phi}$ is canonical, however we would be working with an 
accuracy ${\rm O}(H_{in}/p_{in})$ rather than ${\rm O}(H_{in}^2/p_{in}^2)$. 
This situation is more difficult in terms of the numerical computation
needed, since to achieve the desired accuracy, one has to start the numerical evolution much deeper
inside the horizon. Moreover, initial conditions in the standard isotropic
case are also given to an accuracy of ${\rm O}(H_{in}/p_{in})^2$, therefore we choose
to perform the field rotation to eliminate the ${\bf X} \sim {\rm O}(H)$ terms
and work with the fields ${\bf \Psi}$. The field rotation can be performed trivially
for the case of $2d$ vector perturbations. The case for the $2d$ scalar modes
are more subtle, therefore we leave the discussion of scalar modes to a separate 
publication. 

The canonical quantization procedure outlined above determines the
adiabatic initial conditions for the modes ${\bf \Psi}$. We are
however interested in the values of the entries of ${\bf \Phi}$
evaluated at the end of inflation, which are related to the
primordial power spectra. We will show in the next section that
the matrix ${\bf X}$ along with the off-diagonal entries of
$\omega^2$ is proportional to the anisotropy, so they vanish
before the end of inflation in the two field model. Therefore, it
is possible to choose the transformation matrix ${\bf U}$ that
satisfies ${\bf U}(t_{end})={\bf 1}$, which in turn implies that
${\bf U}^{\dag}\, \left( \omega^2 + {\bf X}^{\dag}\, {\bf X}
\right)\, {\bf U}$ is also diagonal at $t_{end}$ ($t_{end}$ is the
time when inflation ends). Then we have ${\bf\Phi}(t_{end}) =
{\bf\Psi}(t_{end})$ and we use the following equality to compute
the two point functions for the starting modes ${\bf\Phi}$;
\begin{equation}
< \Psi_a^{\dag}\, \Psi_b >|_{t_{end}} = < \Phi_a^{\dag}\, \Phi_b
>|_{t_{end}}
\end{equation}
for the components of the fields. Therefore, instead of working
with the original fields ${\bf \Phi}$ we work with the transformed
fields ${\bf\Psi}$. Notice that passing to the fields ${\bf\Psi}$
was an important step, since the action is in its canonical form
with diagonal mass terms $p^2$ (up to accuracy ${\rm O}(H^2)$) in
the deep UV regime. Therefore, this field can be quantized and $p$
represents the energy of the each particle created by
$\hat{a}^{\dag}(k)$. The quantization procedure also determines
the mode functions $v$, which in turn determines initial
conditions for the perturbations.

In the next subsection, we will discuss the evolution of the $2d$
vector modes, which will be related to the power spectrum of the
gravitational wave polarization $h_{\times}$.

\subsection{2d Vector Perturbations}
\label{sub:2dV}

We now discuss the $2d$ vector perturbations of the two field
inflationary background solution discussed in the previous
section. As we have argued earlier, the quadratic action for the
$2d$ vector and scalar modes decouple, so we concentrate on the
vector piece here. The action for the $2d$ vector perturbations in
momentum space, up to a total time derivative, is calculated as
\begin{eqnarray}
S_{2dV} &=& \frac{M_p^2}{4}\, \int\, dt\, d^3 k\, e^{3 \alpha-2
\sigma}\, {\cal L}_{2dV} \nonumber\\
{\cal L}_{2dV} &=& e^{2 \alpha - 2 \sigma}\, p_L^2\, \left\vert
\dot{\tilde{B}}_3 \right\vert^2 + 2\, e^{-2\alpha}\,
\frac{f^2(\phi)}{M_p^2}\, \left\vert \dot{\alpha}_3 \right\vert^2
- e^{\alpha}\, p_L^2\, \left( \dot{\tilde{B}}_3^*\, B_3 + {\rm
h.c.} \right) + 2\, e^{-\sigma}\, \frac{\tilde{p}_A\,
p_L}{M_p^2}\, \left( i\, \dot{\alpha}_3^*\, \tilde{B}_3 + {\rm
h.c.} \right) \nonumber\\
&& - e^{2\alpha - 2\sigma}\, p_L^2\, \left( p_T^2 - 9 \dot\sigma^2
- \frac{\tilde{p}_A^2}{M_p^2\, f^2(\phi)} \right)\, \left\vert
\tilde{B}_3 \right\vert^2 + 3\, e^{\alpha}\, \dot\sigma\, p_L^2\,
\left( \tilde{B}_3^*\, B_3 + {\rm h.c.} \right) \nonumber\\
&& -2 e^{-2\alpha}\, \frac{p^2}{M_p^2}\, f^2(\phi)\, \left\vert
\alpha_3 \right\vert^2 + 2\, e^{-\alpha + \sigma}\,
\frac{\tilde{p}_A\, p_L}{M_p^2}\, \left( i\, \alpha_3^*\, B_3 +
{\rm h.c.} \right) + e^{2\sigma}\, p^2\, \left\vert B_3
\right\vert^2 \label{action2dv}
\end{eqnarray}
where we have used the physical momenta $p_L \equiv k_L/a$,
$p_T=k_T/b$ and $p=\sqrt{p_L^2+p_T^2}$. As can be seen from
(\ref{action2dv}), the mode $B_3$ is nondynamical, so it can be
integrated out from the action by solving its equation of motion.
The solution for $B_3$ is obtained as
\begin{equation}
B_3 = e^{\alpha - 2 \sigma}\, \left[ \frac{2\, i\, p_L\,
\tilde{p}_A}{M_p^2\, p^2}\, e^{-2\alpha + \sigma}\, \alpha_3 +
\frac{p_L^2}{p^2}\, \left( \dot{\tilde{B}}_3 - 3\, \dot\sigma\,
\tilde{B}_3 \right) \right] \label{solB3}
\end{equation}
We insert the solution (\ref{solB3}) back into the action
(\ref{action2dv}) and up to a total time derivative, we obtain
\begin{eqnarray}
S_{2dV} &=& \frac{1}{2}\, \int\, dt\, d^3k\, \Bigg\{ \vert
\dot{H}_{\times} \vert^2 + \vert \dot{\Delta}_{-} \vert^2 +
\frac{\tilde{p}_A\, p_T}{\sqrt{2}\, M_p\, p\, f(\phi)}\, \left(
i\, \dot{H}_{\times}^*\, \Delta_{-} + i\, H_{\times}\,
\dot{\Delta}_{-}^* + {\rm h.c.} \right) \nonumber\\
&& \qquad\qquad\qquad\qquad -\left( H_{\times}^* \,\,\,\,
\Delta_{-}^*\right)\, \left( \begin{array}{cc} \Omega_{11}^2 & \Omega_{12}^2 \\
\Omega_{12}^{*2} & \Omega_{22}^2 \end{array} \right)\, \left(
\begin{array}{c} H_{\times} \\ \Delta_{-} \end{array} \right)
\Bigg\} \label{action2dvcan}
\end{eqnarray}
where the canonical modes $H_{\times}$ and $\Delta_{-}$ are
related to the starting modes as
\begin{equation}
\tilde{B}_3 = \frac{\sqrt{2}\, p}{p_L\, p_T\, M_p}\,
e^{-\frac{5}{2}\, \alpha + 2 \sigma}\, H_{\times} \,\,\,\, ,
\,\,\,\, \alpha_3 = \frac{e^{-\alpha/2 + \sigma}}{f(\phi)}\,
\Delta_{-} \label{defncan}
\end{equation}
and the mass terms $\Omega^2$ are defined as
\begin{eqnarray}
\Omega_{11}^2 &\equiv& p^2 - \frac{9}{4}\, \dot\alpha^2 +
\frac{3}{4 M_p^2}\, \left( \dot\phi^2 + \dot\phi_1^2\right) +
\frac{9}{2}\, \left( \frac{p_L^4}{p^4} + 6 \frac{p_L^2\,
p_T^2}{p^4} - \frac{p_T^4}{p^4} \right)\, \dot\sigma^2 +
\frac{\tilde{p}_A^2}{2 M_p^2\, f^2(\phi)}\, \left(
\frac{p_L^4}{p^4} - \frac{p_T^4}{p^4} \right) \nonumber\\
\Omega_{12}^2 &\equiv& i\, {\bar\omega}_{12}^2 = \frac{i\,
\tilde{p}_A\, p_T}{\sqrt{2}\, M_p\, p\, f(\phi)}\, \left( \frac{4
p_T^2 - 5 p_L^2}{p^2}\, \dot\sigma + \dot\alpha -
\frac{f'(\phi)}{f(\phi)}\,
\dot\phi \right) \nonumber\\
\Omega_{22}^2 &=& p^2 - \frac{1}{4}\, \dot\alpha^2 + \frac{1}{4
M_p^2}\, \left( \dot\phi^2 + \dot\phi_1^2 \right) - 2\,
\dot\alpha\, \dot\sigma + \frac{1}{2}\, \dot\sigma^2 +
\frac{\tilde{p}_A^2}{2 M_p^2\, f^2(\phi)}\, \frac{5 p_L^2 +
p_T^2}{p^2} - \frac{\tilde{p}_A^2}{f^2(\phi)}\,
\frac{f'(\phi)^2}{f^2(\phi)} \nonumber\\
&& + \left[ 2 \left( \dot\alpha + \dot\sigma \right)\, \dot\phi +
V'(\phi) \right]\, \frac{f'(\phi)}{f(\phi)} -
\frac{f''(\phi)}{f(\phi)}\, \dot\phi^2
\end{eqnarray}
The action (\ref{action2dvcan}) has regular kinetic terms, and therefore
there is no ghost instability in the $2d$ vector sector, as expected.
Note that the matrix ${\bf X}$ along with the mass matrix terms
$\Omega^2$ satisfy
\begin{eqnarray}
&& \Omega_{12}^2 = {\rm O}\left(H^2\right) \,\,\,\, , \,\,\,
\Omega_{11}^2 = \Omega_{22}^2 = p^2 + {\rm O}\left(H^2\right)
\,\,\, , \nonumber\\
&& \frac{\tilde{p}_A}{M_p\, f(\phi)} = \sqrt{6 \dot\alpha^2 - 6
\dot\sigma^2 - \frac{2}{M_p^2}\, \left( V(\phi) + V_1(\phi_1)
\right) - \frac{1}{M_p^2}\, \left( \dot\phi^2 + \dot\phi_1^2
\right)} \sim {\rm O}(H)
\end{eqnarray}
therefore, the action (\ref{action2dvcan}) is formally of the type
given in (\ref{genericaction}) with ${\bf \Phi}$ and matrix ${\bf
X}$ given by
\begin{equation}
{\bf \Phi} \equiv \left( \begin{array}{c} H_{\times} \\ \Delta_{-} \end{array} \right)
\,\,\,\, , \,\,\,\, {\bf X} \equiv \left( \begin{array}{cc} 0 & \lambda \\
\lambda & 0
\end{array} \right) \,\,\,\,\, , \,\,\,\,\,\, \lambda \equiv
\frac{\tilde{p}_A\, p_T}{\sqrt{2}\, M_p\, f(\phi)\, p}
\end{equation}
We can therefore determine the transformation matrix ${\bf U}$,
that eliminates the mixed terms $\dot{\bf \Phi}^{\dag}\, {\bf X}\,
{\bf \Phi}$ by solving $\dot{\bf U} + i\, {\bf X}\, {\bf U} = 0$,
which can be solved by diagonalizing the matrix ${\bf X}$. Notice
that ${\bf R}^{\dag}\, {\bf X}\, {\bf R} = D_X$ where
\begin{equation}
{\bf R} = \frac{1}{\sqrt{2}}\, \left( \begin{array}{cc} 1 & -1 \\
1 & 1 \end{array} \right) \,\,\,\, , \,\,\,\, D_X = \left(
\begin{array}{cc} \lambda & 0 \\ 0 & -\lambda \end{array} \right)
\end{equation}
Therefore (\ref{eqnU}) is solved by
\begin{equation}
{\bf U} = {\bf R}\, e^{-i\, \int^t\, D_X\, dt'}\, {\bf W}_0
\label{solnU-2dV}
\end{equation}
where ${\bf W}_0$ is a constant unitary matrix, which does not
have any physical relevance. Therefore, we choose ${\bf W}_0 =
{\bf R}^{\dag}$, so that ${\bf U} = {\bf 1}$ when $\lambda=0$
(i.e. when anisotropy vanishes). We can now determine the new mass
matrix for the transformed fields ${\bf \Psi}$, which are given by
\begin{equation}
\Omega_{\Psi}^2 = {\bf U}^{\dag}\, \left[ \Omega^2 + {\bf
X}^{\dag}\, {\bf X} \right]\, {\bf U}
\end{equation}
where the entries of $\Omega_{\Psi}^2$ are explicitly given by
\begin{eqnarray}
\Omega_{\Psi\, 11}^2 &=& \lambda^2 + \frac{\Omega_{11}^2 +
\Omega_{22}^2}{2} + \frac{\Omega_{11}^2 - \Omega_{22}^2}{2}\,
\cos\left[ 2 {\cal I}(t) \right] + {\bar\omega}_{12}^2\,
\sin\left[ 2 {\cal I}(t) \right] \nonumber\\
\Omega_{\Psi\, 22}^2 &=& \lambda^2 + \frac{\Omega_{11}^2 +
\Omega_{22}^2}{2} - \frac{\Omega_{11}^2 - \Omega_{22}^2}{2}\,
\cos\left[ 2 {\cal I}(t) \right] - {\bar\omega}_{12}^2\,
\sin\left[ 2 {\cal I}(t) \right] \nonumber\\
\Omega_{\Psi\, 12}^2 &=& -\Omega_{\Psi\, 21}^2 = i\, \left\{
{\bar\omega}_{12}^2\, \cos\left[ 2 {\cal I}(t) \right] -
\frac{\Omega_{11}^2 - \Omega_{22}^2}{2}\, \sin\left[ 2 {\cal I}(t)
\right] \right\} \label{om2new}
\end{eqnarray}
where
\begin{equation}
{\cal I}(t) \equiv \int_{t_k}^{t}\, \lambda(t')\, dt' = \Bigg\{
\begin{array}{c} {\rm constant} \,\,\,\,\,\,\,\,\,\, t > t_{iso} \\ \int_{t_k}^{t}\, \lambda(t')\, dt'
\,\,\,\,\,  t < t_{iso} \end{array} \label{integral}
\end{equation}
Here, $t_{iso}$ is the time when the universe isotropizes so
$\lambda(t \ge t_{iso})=0$. The lower limit $t_k$ in the integral
is chosen for each comoving momentum $k$ such that the constant
value for $t>t_{iso}$ is unity~\footnote{It is possible to
keep the lower limit in the integral arbitrary, since it has no
physical relevance. However, for an arbitrary value, the mass terms for the fields ${\bf \Psi}$
will in general be nondiagonal at the end of inflation. In this case, one has to
further rotate the fields (by a constant matrix) to obtain the physical modes. We perform
this step from the beginning by choosing $t_k$ corrrectly, so that ${\bf U}=1$ at
the end of inflation.}. This ensures that ${\bf
U}(t_{iso}) = {\bf U}(t_{end}) = {\bf 1}$ since
\begin{equation}
{\bf U} \equiv \left( \begin{array}{cc} \cos\left[ {\cal I}(t)
\right] & -i \sin\left[ {\cal I}(t) \right] \\ -i \sin\left[ {\cal
I}(t) \right] & \cos\left[ {\cal I}(t) \right] \end{array} \right)
\rightarrow {\bf 1} \,\,\,\,\, {\rm when}\,\,\,\,\, t \ge t_{iso}
\end{equation}
Finally, the fields ${\bf\Psi}$ satisfy the evolution equations
\begin{equation}
\ddot{\Psi}_a + \Omega_{\Psi\, ab}^2\, \Psi_b = 0 \label{eqnPsi}
\end{equation}
with initial conditions on the entries of $\Psi_a$ set by the form
of the mode function $v(t)$ which are given deep inside the
horizon as
\begin{eqnarray}
\Psi_{a\, in} &=& \frac{1}{\sqrt{2p}} + {\rm O}(H^2)
\nonumber\\
\dot{\Psi}_{a\, in} &=& \frac{1}{\sqrt{2p}}\, \left( -i\, p -
\frac{\dot{p}}{2 p} \right) + {\rm O}(H^2) \label{initPsi}
\end{eqnarray}
up to a nonphysical phase.

In the next section, we will numerically evaluate the evolution
equations (\ref{eqnPsi}) with adiabatic initial conditions
(\ref{initPsi}) for a range of comoving momenta, and determine the
two point functions $< \Psi_{a}^{\dag}\, \Psi_{b} >|_{t_{end}}$
which coincides with $< \Phi_{a}^{\dag}\, \Phi_{b} >|_{t_{end}}$
as we argued before. Finally, using (\ref{defncan}) and
(\ref{long-2d-metric}) we deduce the power spectrum (which is
related to the calculated two point functions) for the
gravitational wave polarization $h_{\times}$. We will provide the
study of the $2d$ scalar perturbations in the appendix and leave
the study of the power spectrum for the scalar modes and the comparison
with observations in a separate publication.

\section{Power Spectrum}
\label{sec:power}

\begin{figure}[h]
\centerline{
\includegraphics[width=0.5\textwidth]{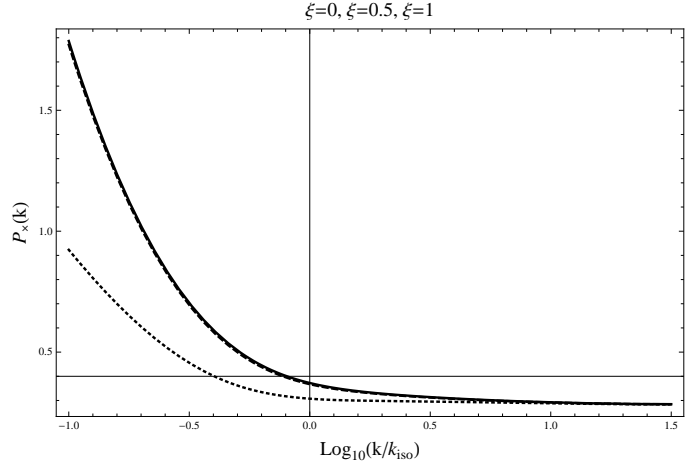}
} \caption{Parts of the power spectrum for $h_{\times}$ are shown
for $\xi=0$ (straight line), $\xi=0.5$ (dashed line) and
$\xi=1$(dotted line).} \label{fig:plot5}
\end{figure}

\begin{figure}[h]
\centerline{
\includegraphics[width=0.4\textwidth]{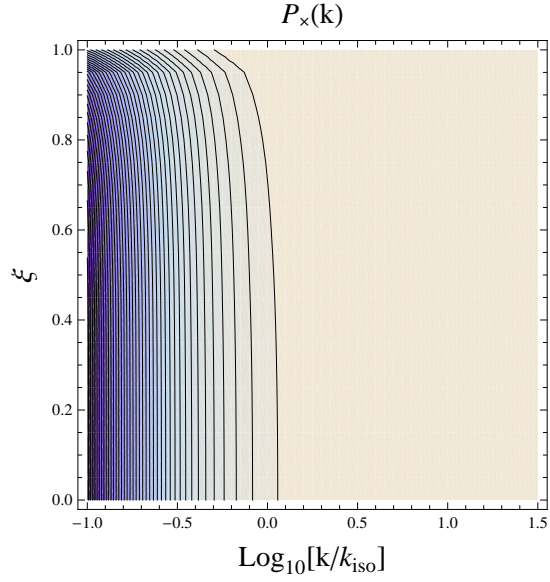}
} \caption{The contour plot of the full power spectrum
$P_{\times}(k)$.} \label{fig:plot6}
\end{figure}

\begin{figure}[h]
\centerline{
\includegraphics[width=0.5\textwidth]{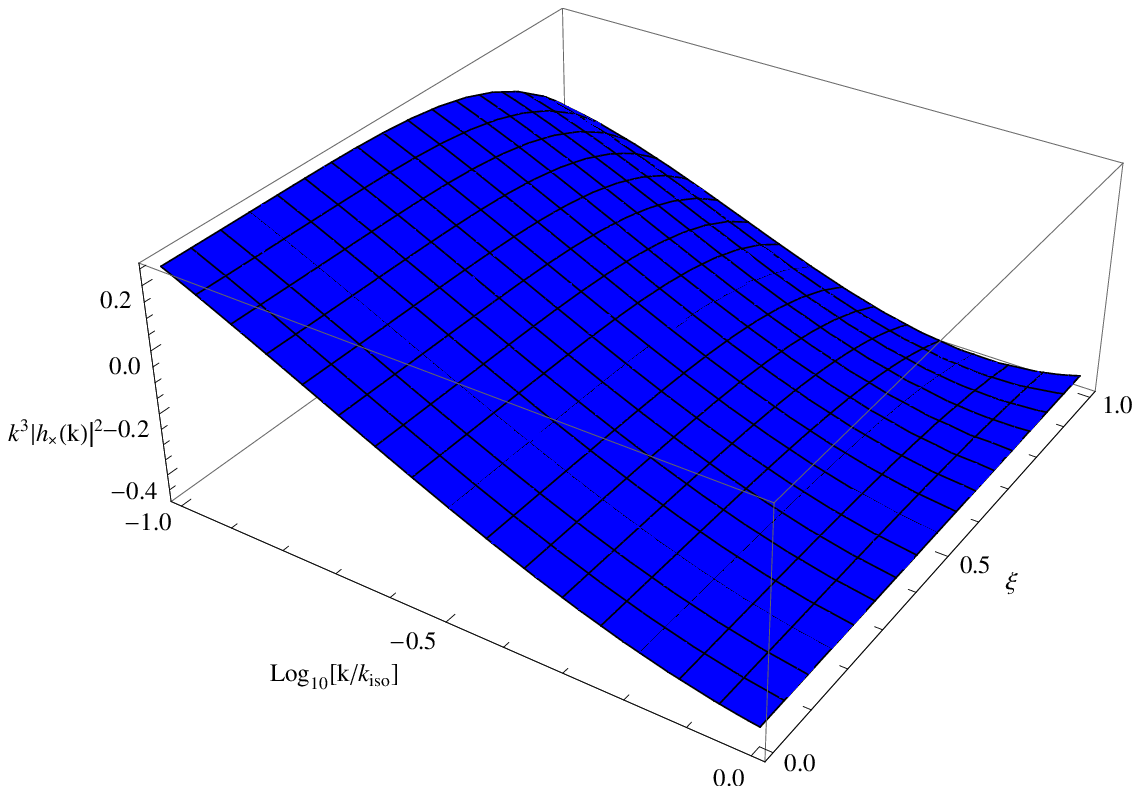}
\includegraphics[width=0.5\textwidth]{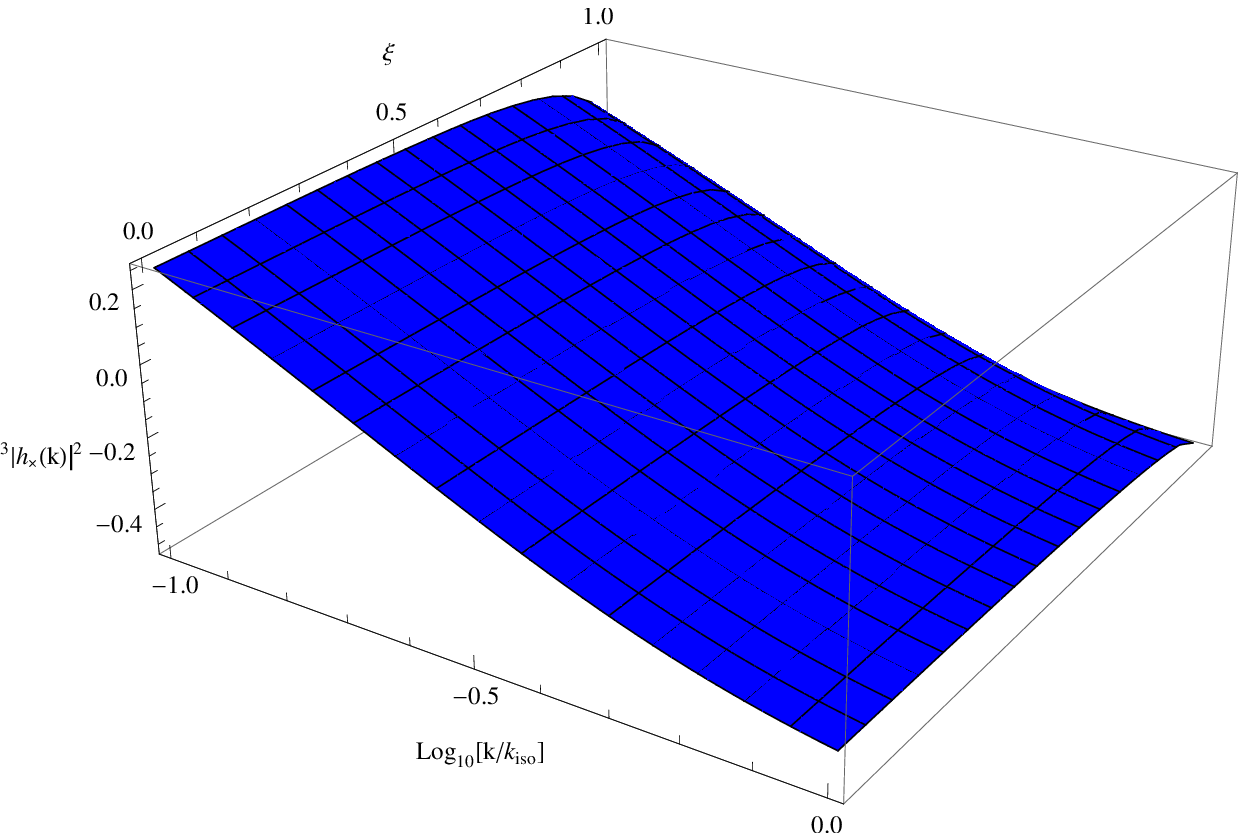}
} \caption{The left panel shows the fitted spectrum using (\ref{powerexp}) and
the right panel the numerical spectrum for $k<k_{iso}$. The fitted and numerical
spectra are in good agreement.} \label{fig:plot7}.
\end{figure}

\begin{figure}[h]
\centerline{
\includegraphics[width=0.5\textwidth]{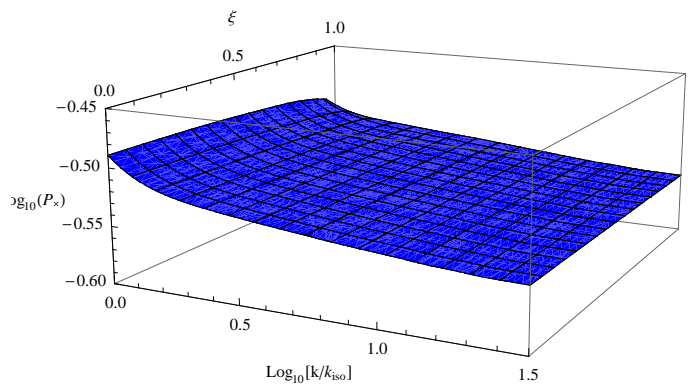}
} \caption{The power spectra at small scales for $k>k_{iso}$,
which is of the standard nearly scale invariant form $P_{\times}
\propto k^{n_T}$.} \label{fig:plot8}
\end{figure}

This section is devoted to the study of the power spectrum of the
gravitational wave polarization $h_{\times}$. We will use the
results of the previous sections to compute the evolution of the
fields $\Psi$ using their equations of motion given in
(\ref{eqnPsi}) with the elements of the mass matrices given in
(\ref{om2new}). We perform the integration for a range of
co-moving momenta $k_L$ and $k_T$, which we parameterize as
\begin{equation}
k_L = \xi\, k \,\,\,\,\, , \,\,\,\,\, k_T = \sqrt{1-\xi^2}\, k
\,\,\,\,\,\, , \,\,\,\,\,\, k \equiv \sqrt{k_L^2 + k_T^2}
\,\,\,\,\,\, , \,\,\,\,\, 0 \le \xi \le 1.
\end{equation}
Therefore, $\xi$ is the cosine of the angle between the comoving
momentum and the longitudinal component $k_L$. For each comoving
mometa labelled by the two numbers $\{k, \xi\}$, initial
conditions are given in equation (\ref{initPsi}). The initial time
$t_{in}$ is chosen to be the moment when
\begin{equation}
\frac{k}{a(t_{in})} = 50\, H(t_{in}) \,\,\,\,\, {\rm or}
\,\,\,\,\, k = 50\, \dot{\alpha}(t_{in})\, e^{\alpha(t_{in})}
\end{equation}
so that the physical momentum $p$ is sufficiently larger than the
Hubble scale and the adiabatic initial conditions (\ref{initPsi})
are accurate. Since the mass matrix $\Omega_{\Psi}^2$ depends on
the background quantities, we integrate the equations
(\ref{eqnPsi}) together with the background equations
(\ref{evolution-2}) with initial conditions and parameters $m_1,
\, m_2$ set as explained in subsection (\ref{sub:2field}). Using
(\ref{long-2d-metric}) and (\ref{defncan}) we can relate the
gravitational wave polarization $h_{\times}$ to the canonical
field $H_{\times}$ at the end of inflation (when
$\sigma=\dot\sigma=0$, so $a=b$)
\begin{equation}
h_{\times} = \frac{\sqrt{2}}{i\, M_p}\, e^{-\frac{3}{2}\,
\alpha}\, H_{\times}. \label{htoH}
\end{equation}
We define the power spectrum for the gravitational wave mode
$h_{\times}$ as
\begin{equation}
P_{\times}(\overrightarrow{k}) \equiv \left( k^3\, \vert
h_{\times} \vert^2 \right)|_{t=t_{end}}
\end{equation}
At the end of inflation, the fields $\Psi$ is identical to the
starting fields $\Phi$, so the power spectrum of the gravitational
wave mode $h_{\times}$is obtained by evaluating the correlation
function $< \Psi_1^*\, \Psi_1 > = < \Phi_1^*\, \Phi_1 > $ and
using (\ref{htoH}) which gives
\begin{equation}
P_{\times}(\overrightarrow{k}) = \frac{2 k^3}{M_p^2}\, e^{-3\,
\alpha(t_{end})}\, \vert \Psi_1(t_{end}) \vert^2.
\end{equation}
In Figures~\ref{fig:plot5},\ref{fig:plot6} we show the power
spectrum obtained from the numerical integration of the background
and perturbation equations. In Fig~\ref{fig:plot5} , parts of the
power spectrum is shown for fixed values of $\xi = 0, \, 0.5, \,
1$ and $10^{-1}\, k_{iso} \le k \le 10^{3/2}\, k_{iso}$. We have
define $k_{iso}$ to be the value of the comoving momentum, when
$k=a(t_{iso})\, H(t_{iso}) = \dot{\alpha}(t_{iso})\,
e^{\alpha(t_{iso})}$ where $t_{iso}$ is the time of isotropization
(we define it to be the time when $h/H=10^{-3}$ numerically). In
Fig.~\ref{fig:plot6}, we show contour plot of the full spectrum
$P_{\times}(\overrightarrow{k})$.
At large scales for $k>k_{iso}$, there is more power and the
spectrum has angular dependence. This in turn could be related to
the CMB anomalies. Note that the spectrum reduces to the standard
nearly scale invariant form when $k > k_{iso}$. Indeed, we have
checked that for $k > k_{iso}$, the power spectrum reduces to
\begin{equation}
P_{\times}(k > k_{iso}) = A_T\, k^{n_T} \,\,\,\,\, , \,\,\,\,
n_{T} \equiv -2\, \epsilon = -\frac{\dot\phi_1^2}{\dot{\alpha}^2\,
M_p^2}
\end{equation}
where $\epsilon$ is the standard slow-roll parameter for the field
$\phi_1$. Indeed, the the spectrum can be written as
\begin{equation}
P_{\times}(\overrightarrow{k}) = \Bigg\{ \begin{array}{c} a_T\,
\left( \frac{k}{k_{iso}} \right)^{f(\xi, \, k)} \,\,\,\,\,\, k \la k_{iso} \\
a_T\, \left( \frac{k}{k_{iso}} \right)^{n_T} \,\,\,\,\,\,\,\,\, k
\ga k_{iso} \end{array}. \label{Ptimes}
\end{equation}
The form of the function $f(\xi, k)$ is unknown, but it is
possible to consider a generic expansion of the form
\begin{eqnarray}
&& {\rm Log}_{10}P_{\times} = {\rm Log}_{10}a_T + \left( c_1 +
d_1\, \xi^{n_1} \right)\, x + \left( c_2 + d_2\, \xi^{n_2}
\right)\, x^2 + \left( c_3 + d_3\, \xi^{n_3} \right)\, x^3 + \dots
\nonumber\\
&& x \equiv {\rm Log}_{10}\left( \frac{k}{k_{iso}} \right)
\label{powerexp}
\end{eqnarray}
We fix the unknown parameters ${\rm Log}_{10}a_T$ (which coincides
with the ${\rm Log}_{10}a_T$ of the spectrum for $k<k_{iso}$) and
$c_1,\, c_2, \, c_3$ by fitting to the numerical spectrum when
$\xi=0$. Then, we fix the parameters $d_1,\, d_2, \, d_3$ by
fitting to the numerical spectrum when $\xi=1$ (using the previous
values of ${\rm Log}_{10}a_T$ and $c_1, \, c_2, \, c_3$). Finally,
we obtain the values of $n_1, \, n_2, \, n_3$ by fitting to a
spectrum at an intermediate value of $\xi$. Increasing the number
of terms in the expansion (\ref{powerexp}), increases the accuracy
of the fit. For the spectra we have obtained above, the unknown
parameters are obtained as
\begin{eqnarray}
&& {\rm Log}_{10}a_T \simeq -0.43, \,\,\, c_1 \simeq -0.23, \,\,\,
c_2 \simeq 0.83, \,\,\, c_3 \simeq 0.38 \nonumber\\
&& d_1 \simeq 0.82, \,\,\, d_2 \simeq 1.21, \,\,\, d_3 \simeq 0.69
\nonumber\\
&& n_1 \simeq 3.94, \,\,\, n_2 \simeq 3.38, \,\,\, n_3 \simeq 3.34
\end{eqnarray}
In Fig.~\ref{fig:plot7}, we show the fitted and numerical power
spectra for $k< k_{iso}$, which are in good agreement. In Fig.~\ref{fig:plot8}, 
we show the numerical spectra for $k> k_{iso}$, which is of the standard nearly scale invariant form.
The power spectrum for the gravitational wave mode $h_{\times}$ is
not of the form $P_{\times} = P_{\times, \, iso}(k) + \delta
P_{\times}(k,\xi)$ as one would naively expect for small
anisotropy for $k < k_{iso}$, with $\vert \delta P_{\times} \vert
\ll \vert P_{\times, \, iso} \vert$. This is due to the fact the
the anisotropic background attractor is not continuously connected
to standard isotropic solutions, i.e. there is no parameter with a
certain limit which takes the anisotropic solution to the
isotropic one. This is in contrast for example to the ACW
model~\cite{acw}, where in the limit $m^2 \rightarrow 0$ ($m^2$ is
the norm of the VEV of the vector field in the model), the
background solution reduces to an isotropic limit. For the current
case, attractors with $c>1$ (anisotropic) and $c < 1$ (isotropic)
are disconnected in the limit $c\rightarrow 1$. Therefore, the
behavior of the perturbations are dramatically different for $k <
k_{iso}$ and the spectrum is largely enhanced in this region
($\vert \delta P_{\times} \vert \ll \vert P_{\times, \, iso}
\vert$ is not satisfied). For phenomenological considerations,
this result is not a desired one, since the CMB data would be
consistent with a statistical anisotropy at large scales that
would satisfy $\vert \delta P_{\times} \vert \ll \vert P_{\times,
\, iso} \vert$. However, this model is the only vector field
driven anisotropic inflationary model that does not suffer from
any instability to our knowledge. Therefore, we believe that the
results obtained in this section are still interesting.

\section{Conclusions}
\label{sec:conclusions}

In this work we have considered anisotropic inflationary models
with coupled vector and scalar fields. The coupling of the scalar
field to the vector field is chosen to preserve the $U(1)$ gauge
symmetry of the Lagrangian, therefore, the longitudinal
polarization of the vector field, which has been shown to cause 
instabilities~\cite{hcp1,hcp2,hcp3}, does not exist. The
anisotropic expansion is achieved when the vector field has a VEV
along one of the spatial directions. The anisotropic solution is
an attractor, with anisotropy increasing towards the end of
inflation, so these types of models are counter examples to the
cosmic no hair conjecture~\cite{Wald}. We have considered a generalization of
the original model introduced in reference~\cite{Watanabe:2009ct},
to include additional scalar fields, and specifically concentrated
on the double scalar field case. This modification was done in
order to achieve isotropization of the universe before the end of
inflation. The original model with a single scalar field coupled
to the vector field has an anisotropic inflationary background
which persists until the end of inflation. In this type of
background, perturbations at all scales are modified. In the two field
modification, the extra field is not coupled to the vector field,
and it is responsible for the overall isotropic expansion of the
universe only. We have chosen the ratio of the masses of the
scalar fields such that the inflationary expansion takes place in
two steps, the first being anisotropic and the second is
isotropic. Therefore, only the largest scale perturbations are
modified and at small scales the standard spectrum is recovered.
As we have discussed in the introduction, vector fields
with a nonvanishing VEV have been introduced since isotropization takes place
very quickly in a Bianchi-I background, (when anisotropy is only coming from initial conditions)
and thus leads to a fine tuning. We have introduced a two phase inflation, with the second phase
being isotropic. The transition to the second isotropic phase is indeed very quick, 
however the fine tuning has been relaxed compared to the previous case. The second isotropic phase
should still be tuned to last around $60$ e-folds, but this tuning is not so strict, since it only 
affects the scale where the power spectrum becomes statistically isotropic. Previously, the tuning
was more stringent, since inflation must last for $60$ e-folds in order that the largest scales
are modified from the initial anisotropic conditions.  A more important drawback of the inlfationary
setup with only initial anisotropic conditions is that the initial singularity is very close to the 
time of isotropization, which is around one Hubble time. Thus, a calculation of the power spectrum
shows that it becomes divergent at the largest scales~\cite{gcp2}~\footnote{This does not mean that
the model is unstable, it reflects that nonlinear effects become important, and one has to go beyond
linear perturbation theory.}.

We have studied the perturbations around the background
configuration in detail, taking into account all the degrees of
freedom of the system. Following the formalism developed in
references~\cite{gcp,gcp2}, we have classified decoupled sets of
perturbations around the anisotropic background and studied them
separately. Furthermore, we have computed the quadratic action for
perturbations and showed that the model is stable and consistent.
We also found linear combinations of the perturbations
which canonically normalize the action. This enabled us to
quantize the system which in turn determined the initial
conditions for the perturbations in the adiabatic vacuum. We then
numerically integrated the equations of motion for the $2d$ vector
perturbations for a range of comoving momenta, which resulted in
the power spectrum of the gravitational wave polarization
$h_{\times}$. We have only computed the quadratic action and
determined the canonical modes for the $2d$ scalar perturbations.
We leave the study of the power spectra and the consequent phenomenology, which in turn is related
to the curvature and $h_+$ gravitational wave spectrum, to a later
publication.

The computed spectrum for $h_{\times}$ has certain interesting
features, which might have some relevance to the observed CMB
anomalies. First of all, the spectrum has angular dependence for
$k<k_{iso}$ which breaks the statistical isotropy at large scales,
therefore leads to an alignment of the lowest multipoles as
described in~\cite{gcp,gcp2,acw}. Moreover, the spectrum reduces
to the standard nearly scale invariant form for $k>k_{iso}$, so
the higher multipoles are not affected from the vector field. We
have shown that the power spectrum is formally of the form
$P_{\times}(\overrightarrow{k}) = P_{\times,\, iso}(k) + \delta
P_{\times}(k,\xi)$.  We expect that this will be true also for the
spectrum of scalar modes. We have also provided a functional form for
$\delta P_{\times}$ by fitting to the numerical spectrum in
equation (\ref{powerexp}). Phenomenologically acceptable
modifications to the large scale spectrum should satisfy $\vert
\delta P \vert \ll \vert P_{iso} \vert$,
however, we have shown that for the model under discussion, this
is not the case. Although our results are obtained for the gravitational
wave mode $h_{\times}$, we expect a similar behavior also for the curvature
spectrum, which can be compared with observations.
At larger scales, the power spectrum is greatly
enhanced satisfying $\vert \delta P_{\times} \vert > \vert
P_{\times, \, iso} \vert$. Although the anisotropy in the
background is small (in the level of a percent, as can be seen
from equation(\ref{sigmadot})), the modification to the spectrum
at large scales is not small. This is due to the fact that the
background attractor solution is disconnected from the standard
isotropic solution and therefore, the behavior of perturbations
are dramatically different. There is no parameter in the attractor
solution that connects it continuously to the isotropic FRW limit.
The anisotropy is always small and proportional to $1/c^2$ and
there is no finite limit in $c$ which can continuously deform the
attractor solution to the isotropic solution. Therefore, the power
spectrum cannot be approximated as to be the isotropic piece plus
a small correction from anisotropy. This issue unfortunately
limits the phenomenology of the model.

Even though the difficulties involved, anisotropic backgrounds
still have potentially interesting phenomenological consequences.
Off-diagonal correlations of the $a_{lm}$ coefficients of CMB
temperature fluctuations arise in such backgrounds and can have
relevance to the alignment of lowest multipoles. Moreover, at the
largest scales, non standard scalar to tensor ratio is obtained,
which can be tested by upcoming CMB experiments. The gravitational
wave modes $h_{+}$ and $h_{\times}$ behave differently at large
scales, therefore a future detection of gravitational waves might
be tested against the consequences of an early stage of
anisotropic inflation. In this paper, we have provided a solid
example of anisotropic inflationary calculation of perturbations
and computed the spectrum of $h_{\times}$ polarization of
gravitational waves. Although the obtained spectrum has
phenomenologically limited implications, we believe that our
results are still relevant, since it is the first complete study
of power spectrum in an anisotropic inflationary model driven by a
vector field to our knowledge. In a further publication, we will
study the spectrum of the curvature ${\cal R}$ perturbation, which
could then be compared with observational data.

\begin{acknowledgments}

The author would like to thank Marco Peloso, Lorenzo Sorbo,
Jose Cembranos for very useful discussions. This work is supported
by the Graduate School at the University of Minnesota under the
Doctoral Dissertation Fellowship.

\end{acknowledgments}

\section{Appendices}
\label{sec:app}

\subsection{Appendix I}
\label{sub:app1}

In this appendix we derive the relations (\ref{long-2d-metric}),
which are used in the main text to compute the power spectrum of
the gravitational wave mode $h_{\times}$. These relations are also
useful when the curvature and $h_{+}$ power spectra are
calculated. For this calculation, one needs to analyze the $2d$
scalar perturbations, which is done in the next appendix. We will
study the power spectrum for the scalar modes in a separate
publication.

We start with the decomposition of perturbations in the
longitudinal gauge, when the universe is isotropic, which is given
by
\begin{equation}
ds^2 = -\left( 1 + 2 \phi \right)\, dt^2 + 2\, a\, V_i\, dt\, dx^i
+ a^2\, \left[ \left( 1 - 2 \psi \right)\, \delta_{ij} + h_{ij}
\right]\, dx^i\, dx^j
\end{equation}
where $i=1,2,3$. $V_i$'s are vector modes satisfying $\partial_i\,
V_i = 0$ and $h_{ij}$'s are transverse-traceless modes satisfying
$\partial_i\, h_{ij} = h_{ii} = 0$. Since the spatial slices are
isotropic, we can fix the comoving momentum to lie entirely along
the x-direction $\overrightarrow{k} = \left( k,\, 0, \, 0\right)$,
without any loss of generality. In this case, the longitudinal
gauge can be expressed in Fourier space as
\begin{equation}
\delta g_{\mu\nu}(k) = \left( \begin{array}{cccc} -2\phi & 0 & V_2
& V_3 \\
 & -2 a^2\, \psi & 0 & 0 \\
 &  & a^2\, \left(-2\, \psi + h_{+}\right) & -a^2\, h_{\times} \\
 &  & -a^2\, h_{\times} & a^2\, \left(-2\, \psi - h_{+} \right)
\end{array} \right) \label{long}
\end{equation}
We can now rotate the momentum from the x-direction, to $x-y$
plane in order to make the comparison with the $2d$ decomposition.
Namely, we perform the following rotation over the comoving
momentum:
\begin{equation}
\left( \begin{array}{c} k_L \\ k_T \end{array} \right) =
\frac{1}{k}\, \left( \begin{array}{cc} k_L & - k_T \\ k_T & k_L
\end{array} \right)\, \left( \begin{array}{c} k \\ 0 \end{array}
\right) \label{jacobian}
\end{equation}
where $k=\sqrt{k_L^2 + k_T^2}$. The matrix appearing in
(\ref{jacobian}) also determines the Jacobian matrix which fixes
the coordinate transformation rule from the metric with
$\overrightarrow{k}=(k,0,0)$ (\ref{long}) to a metric with
$\overrightarrow{k}=(k_L,k_T,0)$. The resulting decomposition will
determine the $2d$-decomposed longitudinal gauge. Using the
appropriate scale factors $a$ and $b$ where necessary to take into
account the anisotropy of the universe, the $2d$ decomposed
longitudinal gauge becomes
\begin{equation}
\delta g_{\mu\nu}^{L}(k) = \left( \begin{array}{cccc} -2\phi & -a
\frac{k_T}{k}\, V_2 & b\, \frac{k_L}{k}\, V_2 & b\, V_3 \\
 & a^2\, \left( -2\psi + \frac{k_T^2}{k^2}\, h_{+} \right) &
 -a\,b\, \frac{k_L\, k_T}{k^2}\, h_{+} & a\, b\, \frac{k_T}{k}\,
 h_{\times} \\
  & & b^2\, \left( -2\psi + \frac{k_L^2}{k^2}\, h_{+} \right) &
  -b^2\, \frac{k_L}{k}\, h_{\times} \\
  & & & b^2\, \left( -2\psi - h_{+} \right) \end{array} \right)
  \label{pert-long-k}
\end{equation}
where the superscript "L" indicates that longitudinal gauge is
used. Comparing (\ref{pert-long-k}) with (\ref{pert-k}), we
identify;
\begin{eqnarray}
&& \Phi_L = \phi \,\,\, , \,\,\, \chi_L = \frac{k_T}{i\, k\,
k_L}\, V_2 \,\,\, , \,\,\, B_L = \frac{k_L}{i\, k\, k_T}\, V_2
\,\,\, , \,\,\, B_{3L} = V_3 \,\,\, , \,\,\, \tilde{B}_{3L} =
\frac{-k_T}{i\, k\, k_L}\, h_{\times} \,\,\, , \,\,\, E_{3L} =
\frac{-k_L}{i\, k\, k_T}\, h_{\times} \nonumber\\
&& \Sigma_{L} = \psi + \frac{1}{2}\, h_{+} \,\,\, , \,\,\,
\tilde{B}_L = \frac{1}{k^2}\, h_{+} \,\,\, , \,\,\, E_L = \frac{2
k_L^2 + k_T^2}{2 k^2\, k_T^2}\, h_{+} \label{long-2d}
\end{eqnarray}
We can now find the infinitesimal transformation between our gauge
choice of $\tilde{B}=\Sigma=E=E_3=0$ and the longitudinal gauge
using
\begin{equation}
\delta g_{\mu\nu}^{(\rm{our})} = \delta g_{\mu\nu}^L -
g_{\mu\nu,\sigma}^{(0)}\, \epsilon^{\sigma} -
g_{\mu\sigma}^{(0)}\, \epsilon^{\sigma}_{,\nu} -
g_{\sigma\nu}^{(0)}\, \epsilon^{\sigma}_{,\mu} \label{gauge-trans}
\end{equation}
where
\begin{equation}
\epsilon^{\mu}(k) = \left( \xi^0_L, \, -i\, k_L\, \xi^1_L, \, -i\,
k_T\, \xi_L, \, \xi^3_L \right)
\end{equation}
Finally, the infinitesimal transformation between the two gauges
are determined by
\begin{eqnarray}
&& \xi^0_L = -\frac{1}{H_b}\, \left( \psi + \frac{1}{2}\, h_{+}
\right) \,\,\, , \,\,\, \xi_L = -\frac{2 k_L^2 + k_T^2}{2 k^2\,
k_T^2}\, h_{+} \nonumber\\
&& \xi^1_L = \frac{b}{a}\, \frac{1}{k^2}\, \left( 1 +
\frac{b}{a}\, \frac{2 k_L^2 + k_T^2}{2 k_T^2} \right)\, h_{+}
\,\,\, , \,\,\, \xi^3_L = \frac{k_L}{i\, k\, k_T}\, h_{\times}
\label{long-2d-xi}
\end{eqnarray}
Using these transformations, we can relate the power spectra
calculated in the $2d$ decomposition with the standard isotropic
definitions of power. For instance, using (\ref{gauge-trans}) and
(\ref{long-2d-xi}) the dynamical modes of the metric perturbations
of the $2d$ decomposition are related to the gravitational wave
modes $h_{+},\, h_{\times}$ and $\psi$ (related to curvature
perurbation) as
\begin{eqnarray}
\tilde{B}_3 &=& \frac{i}{k\, k_L}\, \left( \frac{k_T^2 + b/a\,
k_L^2}{k_T} \right)\, h_{\times} \nonumber\\
\Psi &=& \left( 1 - \frac{H_a}{H_b} \right)\, \psi - \left\{
\frac{H_a}{2 H_b} + \frac{1}{k^2}\, \left[ \frac{1}{2}\, k_T^2 +
\frac{b^3}{a^3}\, k_L^2\, \left( 1 + \frac{b}{a}\, \frac{2
k_L^2+k_T^2}{2 k_T^2} \right)\, \right]\, \right\}\, h_{+}.
\nonumber
\end{eqnarray}
Moreover, using the infinitesimal transformation $\delta\phi =
\delta\phi^L - \dot\phi\, \xi^0_L$, the fluctuation of any scalar
field in our gauge is related to the one in longitudinal gauge as
\begin{equation}
\delta\phi = \delta\phi^L + \frac{\dot\phi}{H_b}\, \left( \psi +
\frac{1}{2}\, h_{+} \right). \nonumber
\end{equation}

\subsection{Appendix II}
\label{sub:app2}

In this appendix, we study the $2d$ scalar perturbations. As we
have done for the case of the $2d$ vector perturbations, we insert
the metric and vector field decompositions (\ref{defn-perts-2d})
and $\phi=\phi_0(t) + \delta\phi, \, \phi_1 = \phi_{1\, 0}(t) +
\delta\phi_1 $ into the action (\ref{action}) (with $a=1$), after
fixing the $U(1)$ and general coordinate invariance as described
previously. The quadratic action in momentum space up to a total
time derivative is given by
\begin{eqnarray}
S_{2dS} &=& \frac{1}{2}\, \int\, dt\, d^3k\, e^{3\alpha}\, {\cal
L}_{2dS} \nonumber\\
{\cal L}_{2dS} &=& \vert \delta\dot\phi \vert^2 + \vert
\delta\dot\phi_1 \vert^2 + e^{-2\alpha + 4 \sigma}\, f^2(\phi)\,
\vert \dot\alpha_1 \vert^2 - \dot\phi\, \left( \delta\dot\phi^*\,
\Psi + {\rm h.c.} \right) - \dot\phi\, \left( \delta\dot\phi^*\,
\Phi + {\rm h.c.} \right) \nonumber\\
&& - \dot\phi_1\, \left( \delta\dot\phi_1^*\, \Psi + {\rm h.c.}
\right) - \dot\phi_1\, \left( \delta\dot\phi_1^*\, \Phi + {\rm
h.c.} \right) - 2 M_p^2\, \left(\dot\alpha + \dot\sigma \right)\,
\left( \dot\Psi^*\, \Phi + {\rm h.c.} \right) \nonumber\\
&& + \frac{2 e^{-\alpha + 2 \sigma}\, \tilde{p}_A\,
f'(\phi)}{f(\phi)}\, \left( \dot\alpha_1^*\, \delta\phi + {\rm
h.c.} \right) + e^{-\alpha + 2 \sigma}\, \tilde{p}_A\, \left(
\dot\alpha_1^*\, \Psi + {\rm h.c.} \right) -  e^{-\alpha + 2
\sigma}\, \tilde{p}_A\, \left( \dot\alpha_1^*\, \Phi + {\rm h.c.}
\right) \nonumber\\
&& + e^{-\alpha + 2 \sigma}\, p_L\, f^2(\phi)\, \left(i\,
\dot\alpha_1^*\, \alpha_0 + {\rm h.c.} \right) - \left[ p^2 +
V''(\phi) - \tilde{p}_A^2\, \left( \frac{f'(\phi)^2}{f(\phi)^4} +
\frac{f''(\phi)}{f(\phi)^3}
\right) \right]\, \vert \delta\phi \vert^2 \nonumber\\
&& - \left( p^2 + V_1''(\phi_1) \right)\, \vert \delta\phi_1
\vert^2 + \left[ \tilde{p}_A^2\, \frac{f'(\phi)}{f^3(\phi)} +
V'(\phi) \right]\, \left( \delta\phi^*\, \Psi + {\rm h.c.}
\right) \nonumber\\
&& - \left[ \tilde{p}_A^2\, \frac{f'(\phi)}{f^3(\phi)} + V'(\phi)
\right]\, \left( \delta\phi^*\, \Phi + {\rm h.c.} \right) +
V_1'(\phi_1)\, \left( \delta\phi_1^*\, \Psi + {\rm h.c.} \right) -
V_1'(\phi_1)\, \left( \delta\phi_1^*\, \Phi + {\rm h.c.}
\right) \nonumber\\
&& - e^{\alpha - 2\sigma}\, p_L^2\, \dot\phi\, \left(
\delta\phi^*\, \chi + {\rm h.c.} \right) - e^{\alpha+\sigma}\,
p_T^2\, \dot\phi\, \left( \delta\phi^*\, B + {\rm h.c.} \right) -
e^{\alpha - 2\sigma}\, p_L^2\, \dot\phi_1\, \left(
\delta\phi_1^*\, \chi + {\rm h.c.} \right) \nonumber\\
&& - e^{\alpha + \sigma}\, p_T^2\, \dot\phi_1\, \left(
\delta\phi_1^*\, B + {\rm h.c.} \right) + \frac{2\, \tilde{p}_A\,
p_L\, f'(\phi)}{f(\phi)}\, \left( i\, \delta\phi^*\, \alpha_0 +
{\rm h.c.} \right) + \frac{\tilde{p}_A^2}{f(\phi)^2}\, \vert \Psi
\vert^2 \nonumber\\
&& - M_p^2\, \left( p_T^2 + \frac{\tilde{p}_A^2}{M_p^2\,
f^2(\phi)} \right)\, \left( \Psi^*\, \Phi + {\rm h.c.} \right) - 3
e^{\alpha + \sigma}\, M_p^2\, p_T^2\, \dot\sigma\, \left( \Psi^*\,
B + {\rm h.c.} \right) + \tilde{p}_A\, p_L\, \left( i\,
\Psi^*\, \alpha_0 + {\rm h.c.} \right) \nonumber\\
&& - e^{-2\alpha + 4 \sigma}\, p_T^2\, f^2(\phi)\, \vert \alpha_1
\vert^2 - e^{3\sigma}\, \tilde{p}_A\, p_T^2\, \left(
\alpha_1^*\, B + {\rm h.c.} \right) \nonumber\\
&& + M_p^2\, \left[ \frac{\tilde{p}_A^2}{M_p^2\, f(\phi)^2} +
\frac{\dot\phi^2}{M_p^2} + \frac{\dot\phi_1^2}{M_p^2} + 6
(\dot\sigma^2 - \dot\alpha^2) \right]\, \vert \Phi \vert^2 + 2
e^{\alpha-2 \sigma}\, M_p^2\, p_L^2\, \left( \dot\alpha +
\dot\sigma \right)\, \left( \Phi^*\, \chi + {\rm h.c.} \right)
\nonumber\\
&& + e^{\alpha + \sigma}\, M_p^2\, p_T^2\, \left( 2\dot\alpha -
\dot\sigma \right)\, \left( \Phi^*\, B + {\rm h.c.} \right) -
\tilde{p}_A\, p_L\, \left( i\, \Phi^*\, \alpha_0 + {\rm h.c.}
\right) + \frac{1}{2}\, e^{2 \alpha - 4 \sigma}\, M_p^2\, p_L^2\,
p_T^2\, \vert \chi \vert^2 \nonumber\\
&& - \frac{1}{2}\, e^{2\alpha - \sigma}\, M_p^2\, p_L^2\, p_T^2\,
\left( \chi^*\, B + {\rm h.c.} \right) + \frac{1}{2}\, e^{2\alpha
+ 2 \sigma}\, M_p^2\, p_L^2\, p_T^2\, \vert B \vert^2 + p^2\,
f^2(\phi)\, \vert \alpha_0 \vert^2 \label{action2ds}
\end{eqnarray}
We can see from the above action that the modes $\{ \Phi, \, \chi,
\, B, \, \alpha_0 \}$ are nondynamical, and can be integrated out.
More precisely, we extremize the action (\ref{action2ds}) with
respect to the nondynamical modes to obtain constraint equations,
which we solve for the values of the nondynamical modes. We insert
the solutions back into the action (\ref{action2ds}) to obtain an
action purely in terms of the dynamical modes $\{ \delta\phi, \,
\delta\phi_1, \, \Psi, \, \alpha_1 \}$ up to a total time
derivative. This action reads
\begin{eqnarray}
S_{2dS} &=& \frac{1}{2}\, \int\, dt\, d^3k\, \Bigg\{ \vert \dot{V}
\vert^2 + \vert \dot{V}_1 \vert^2 + \vert \dot{H}_{+} \vert^2 +
\vert \dot{\Delta}_{+} \vert^2 + {\cal F}\, \left( i\, \dot{V}^*\,
\Delta_{+} + i\, V\, \dot{\Delta}_{+}^* + {\rm
h.c.} \right) \nonumber\\
&& \qquad\qquad\qquad + {\cal G}\, \left( i\, \dot{H}_{+}^*\,
\Delta_{+} + i\, H_{+}\, \dot{\Delta}_{+}^* + {\rm h.c.} \right)
\nonumber\\
&& \qquad\qquad\qquad - \left( V^*\, V_1^*\, H_{+}^*\,
\Delta_{+}^* \right)\, \Omega^2\, \left( \begin{array}{c} V \\
V_1 \\ H_{+} \\ \Delta_{+} \end{array} \right) \Bigg\}
\label{action2dscan}
\end{eqnarray}
where ${\cal F}$ and ${\cal G}$ are given by
\begin{equation}
{\cal F} = - \frac{\tilde{p}_A\, f'(\phi)}{f(\phi)^2}\,
\frac{p_T}{p} \,\,\,\, , \,\,\,\, {\cal G} = -
\frac{\tilde{p}_A}{\sqrt{2}\, M_p\, f(\phi)}\, \frac{p_T}{p}
\label{FandG}
\end{equation}
with $f(\phi_1)={\rm Exp}\left[ \frac{4}{M_p^2}\, \int\,
\frac{V_1(\phi_1)}{V_1'(\phi_1)}\, d\phi_1 \right]$. Notice that
the action (\ref{action2dscan}) has regular kinetic terms, therefore
the $2d$ scalar sector is free from ghost instabilities as in the
case of $2d$ vector modes. The elements of the hermitian matrix $\Omega^2$ are given as
\begin{eqnarray}
\Omega_{11}^2 &=& p^2 - \frac{9}{4}\, \dot\alpha^2 + \frac{15\,
\dot\phi^2}{4 M_p^2} + \frac{3\, \dot\phi_1^2}{4 M_p^2} +
\frac{9}{2}\, \dot\sigma^2 + \frac{2\, \dot\phi_1\,
V'(\phi)}{M_p^2\, \dot\alpha} + V''(\phi) -2\, \frac{p^4}{{\cal
D}^2}\, \frac{\dot\phi^4}{M_p^4} - 2\, \frac{p^4}{{\cal D}^2}\,
\frac{\dot\phi^2\, \dot\phi_1^2}{M_p^4}
\nonumber\\
&& + \frac{2\, \left( p_T^2 - 2 p_L^2 \right)\, V'(\phi)}{{\cal
D}}\, \frac{\dot\phi}{M_p^2\, \dot\alpha}\, \dot\sigma - 3\,
\left( 4\, \frac{\left( 2 p_L^2 - p_T^2 \right)\, p^2}{{\cal
D}^2}\, \dot\alpha + \frac{8 p_L^4 - 8 p_L^2\, p_T^2 + 5
p_T^4}{{\cal D}^2}\, \dot\sigma \right)\,
\frac{\dot\phi^2}{M_p^2}\, \dot\sigma \nonumber\\
&& + \frac{\tilde{p}_A^2}{2 M_p^2\, f(\phi)^2}\, \Big[ 1 + \frac{2
M_p^2\, \left( 3 p_L^2 - p_T^2 \right)\, f'(\phi)^2}{p^2\,
f(\phi)^2} - 4\, \frac{p_T^2\, p^2}{{\cal D}^2}\,
\frac{\dot\phi^2}{M_p^2} - 8\, \frac{p_L^2}{{\cal D}}\,
\frac{f'(\phi)}{f(\phi)}\, \dot\phi - 2 M_p^2\,
\frac{f''(\phi)}{f(\phi)} \Big] \nonumber
\end{eqnarray}
\begin{eqnarray}
{\Omega}_{12}^2 &=& 2\, \dot\sigma\, \frac{\left( 2 p_L^2 - p_T^2
\right)\, p^2}{{\cal D}^2}\, \left[ \frac{\dot\phi_1\,
V'(\phi)}{M_p^2} + \frac{\dot\phi\, V_1'(\phi_1)}{M_p^2} - 6\,
\frac{p_L^4 - p_L^2\, p_T^2 + p_T^4}{\left( 2 p_L^2 -
p_T^2\right)\, p^2}\, \frac{\dot\phi\, \dot\phi_1}{M_p^2}\,
\dot\sigma \right] \nonumber\\
&& - 2\, \frac{p^4}{{\cal D}^2}\, \left[ \frac{\dot\phi^3\,
\dot\phi_1}{M_p^4} - 2\, \frac{\dot\phi_1\, V'(\phi)}{M_p^2}\,
\dot\alpha + \frac{\dot\phi}{M_p}\, \left(
\frac{\dot\phi_1^3}{M_p^3} - 2\, \frac{V_1'(\phi_1)}{M_p}\,
\dot\alpha - 6\, \frac{\dot\phi_1}{M_p}\, \dot\alpha^2 \right)
\right] \nonumber\\
&& - \frac{2\, \tilde{p}_A^2}{M_p^2\, f(\phi)^2}\,
\frac{\dot\phi\, \dot\phi_1}{M_p^2}\, \frac{p_T^2\, p^2}{{\cal
D}^2} - \frac{2\, \tilde{p}_A^2}{M_p^2\, f(\phi)^2}\,
\frac{f'(\phi)}{f(\phi)}\, \dot\phi_1\, \frac{p_L^2}{{\cal D}}
\nonumber
\end{eqnarray}
\begin{eqnarray}
{\Omega}_{13}^2 &=& -\frac{3\, \sqrt{2}\, p_T^2\, p^2\,
\dot\sigma}{{\cal D}^2}\, \left[ \frac{\dot\phi^3}{M_p^3} +
\frac{\dot\phi\, \dot\phi_1^2}{M_p^3} - 6\, \frac{\dot\phi}{M_p}\,
\left( \dot\alpha + \dot\sigma \right)\, \left( \dot\alpha +
\frac{p_L^2 - p_T^2}{p^2}\, \dot\sigma \right) -
\frac{V'(\phi)}{M_p}\, \left( 2 \dot\alpha + \frac{2 p_L^2 -
p_T^2}{p^2}\, \dot\sigma \right) \right] \nonumber\\
&& - \frac{\sqrt{2}\, \tilde{p}_A^2\, p_T^2}{M_p^2\, f(\phi)^2\,
{\cal D}^2}\, \Bigg[ 3 p_T^2\, \frac{\dot\phi}{M_p}\, \dot\sigma +
M_p\, \frac{f'(\phi)}{f(\phi)}\, \Big( 4\, p^2\, \dot\alpha^2 +
2\, \left( 7 p_L^2 - 2 p_T^2 \right)\,
\dot\alpha\, \dot\sigma \nonumber\\
&& \qquad\qquad\qquad\qquad\qquad\qquad\qquad  + \frac{\left( 2
p_L^2 - p_T^2\right)\, \left( 5 p_L^2 - p_T^2\right)}{p^2}\,
\dot\sigma^2 \Big) \Bigg] \nonumber
\end{eqnarray}
\begin{eqnarray}
{\Omega}_{14}^2 &=& \frac{i\, \tilde{p}_A\, p_T}{p\, f(\phi)}\,
\Bigg\{ - \frac{2\, \tilde{p}_A^2\, p^2}{M_p^2\, f(\phi)^2\, {\cal
D}^2}\, \left[ \frac{\dot\phi}{M_p^2}\, p_T^2 +
\frac{f'(\phi)}{f(\phi)}\, p_L^2\, \left( 2 \dot\alpha + \frac{2
p_L^2 - p_T^2}{p^2}\, \dot\sigma \right) \right]
\nonumber\\
&& -2\, \frac{p^4}{{\cal D}^2\, M_p}\, \left[
\frac{\dot\phi^3}{M_p^3} + \frac{\dot\phi\, \dot\phi_1^2}{M_p^3} -
6\, \frac{\dot\phi}{M_p}\, \left( \dot\alpha + \dot\sigma
\right)\, \left( \dot\alpha + \frac{p_L^2-
p_T^2}{p^2}\, \dot\sigma \right) \right] \nonumber\\
&& + 2\, \frac{p^4}{{\cal D}^2\, M_p}\, \frac{V'(\phi)}{M_p}\,
\left( 2\, \dot\alpha + \frac{2 p_L^2 - p_T^2}{p^2}\, \dot\sigma
\right) - \frac{f'(\phi)}{f(\phi)}\, \left( \dot\alpha + \frac{7
p_L^2 - 2 p_T^2}{p^2}\, \dot\sigma \right) + \dot\phi\,
\frac{f''(\phi)}{f(\phi)} \Bigg\} \nonumber
\end{eqnarray}
\begin{eqnarray}
{\Omega}_{22}^2 &=& p^2 - \frac{9}{4}\, \dot\alpha^2 + \frac{3\,
\dot\phi^2}{4 M_p^2} + \frac{15\, \dot\phi_1^2}{4 M_p^2} +
\frac{9}{2}\, \dot\sigma^2 + \frac{2\, \dot\phi_1\,
V_1'(\phi_1)}{M_p^2\, \dot\alpha} + V_1''(\phi_1) - 2\,
\frac{p^4}{{\cal D}^2}\, \frac{\dot\phi_1^4}{M_p^4} - 2\,
\frac{p^4}{{\cal D}^2}\, \frac{\dot\phi^2\, \dot\phi_1^2}{M_p^4}
\nonumber\\
&& + \frac{2\, \left( p_T^2 - 2 p_L^2 \right)\,
V_1'(\phi_1)}{{\cal D}}\, \frac{\dot\phi_1}{M_p^2\, \dot\alpha}\,
\dot\sigma - 3\, \left[ 4\, \frac{\left( 2 p_L^2 - p_T^2\right)\,
p^2}{{\cal D}^2}\, \dot\alpha + \frac{8 p_L^4 - 8\, p_L^2\, p_T^2
+ 5 p_T^4}{{\cal D}^2}\, \dot\sigma \right]\,
\frac{\dot\phi_1^2}{M_p^2}\, \dot\sigma \nonumber\\
&& + \frac{\tilde{p}_A^2}{2 M_p^2\, f(\phi)^2}\, \left( 1 - 4\,
\frac{p_T^2\, p^2}{{\cal D}^2}\, \frac{\dot\phi_1^2}{M_p^2}
\right) \nonumber
\end{eqnarray}
\begin{eqnarray}
\Omega_{23}^2 &=& -\frac{3\, \sqrt{2}\, p_T^2\, p^2\,
\dot\sigma}{{\cal D}^2}\, \Bigg[ \frac{\dot\phi_1^3}{M_p^3} +
\frac{\dot\phi^2\, \dot\phi_1}{M_p^3} - 6\,
\frac{\dot\phi_1}{M_p}\, \left( \dot\alpha + \dot\sigma \right)\,
\left( \dot\alpha + \frac{p_L^2 - p_T^2}{p^2}\, \dot\sigma \right)
- \frac{V_1'(\phi_1)}{M_p}\, \left( 2\, \dot\alpha + \frac{2 p_L^2
- p_T^2}{p^2}\, \dot\sigma \right) \nonumber\\
&& + \frac{\tilde{p}_A^2}{M_p^2\, f(\phi)^2}\,
\frac{\dot\phi_1}{M_p}\, \frac{p_T^2}{p^2} \Bigg] \nonumber
\end{eqnarray}
\begin{eqnarray}
\Omega_{24}^2 &=& -\frac{2\, i\, \tilde{p}_A}{M_p\, f(\phi)}\,
p_T\, p\, \Bigg[ \frac{p^2}{{\cal D}^2}\,
\frac{\dot\phi_1^3}{M_p^3} - \frac{V_1'(\phi_1)}{{\cal D}\, M_p} +
\frac{\tilde{p}_A^2}{M_p^2\, f(\phi)^2}\,
\frac{\dot\phi_1}{M_p}\, \frac{p_T^2}{{\cal D}^2} \nonumber\\
&& \qquad\qquad\qquad\qquad  + \frac{\dot\phi_1}{M_p}\, \left(
\frac{p^2}{{\cal D}^2}\, \frac{\dot\phi^2}{M_p^2} - 6\, \left(
\dot\alpha + \dot\sigma \right)\, \frac{p^2\, \dot\alpha + (p_L^2
- p_T^2)\, \dot\sigma}{{\cal D}^2} \right) \Bigg] \nonumber
\end{eqnarray}
\begin{eqnarray}
\Omega_{33}^2 &=& p^2 - \frac{1}{4}\, \dot\alpha^2 + \frac{3\,
(\dot\phi^2 + \dot\phi_1^2)}{4 M_p^2} - 8\, \frac{p^4}{{\cal
D}^2}\, \dot\alpha^4 - \frac{9 p_T^4\, \dot\sigma^2}{{\cal D}^2}\,
\frac{\dot\phi^2 + \dot\phi_1^2}{M_p^2} - 8\, \frac{\left( 2 p_L^2
- p_T^2 \right)\, p^2}{{\cal D}^2}\, \dot\alpha^3\,
\dot\sigma \nonumber\\
&& + 2\, \frac{5 p_L^2 + 58\, p_L^2\, p_T^2 + 35\, p_T^4}{{\cal
D}^2}\, \dot\alpha^2\, \dot\sigma^2 + 18\, \frac{2 p_L^4 + 9\,
p_L^2\, p_T^2 - p_T^4}{{\cal D}^2}\, \dot\alpha\, \dot\sigma^3 +
\frac{9}{2}\, \frac{4 p_L^4 + 12\, p_L^2\, p_T^2 - 11 p_T^4}{{\cal
D}^2}\, \dot\sigma^4 \nonumber\\
&& + \frac{\tilde{p}_A^2}{2 M_p^2\, f(\phi)^2}\, \frac{p_L^2 -
p_T^2}{p^2} - \frac{9\, \tilde{p}_A^2}{M_p^2\, f(\phi)^2}\,
\frac{p_T^6}{p^2\, {\cal D}^2}\, \dot\sigma^2 \nonumber
\end{eqnarray}
\begin{eqnarray}
\Omega_{34}^2 &=& \frac{i\, \tilde{p}_A\, p_T}{\sqrt{2}\, M_p\,
p\, f(\phi)}\, \Bigg\{ -\frac{6 \tilde{p}_A^2}{M_p^2\,
f(\phi)^2}\, \frac{p_T^4}{{\cal D}^2}\, \dot\sigma +
\frac{f'(\phi)}{f(\phi)}\, \dot\phi - 4\, \frac{p^4}{{\cal
D}^2}\, \dot\alpha^3 \nonumber\\
&& \qquad\qquad\qquad\qquad - 6\, \frac{p_T^2\, p^2}{{\cal D}^2}\,
\left( \frac{\dot\phi^2 + \dot\phi_1^2}{M_p^2} - 2\, \frac{p_L^2 +
4 p_T^2}{p_T^2}\, \dot\alpha^2 \right)\, \dot\sigma + 9\, \frac{4
p_L^2 + 16\, p_L^2\, p_T^2 - p_T^4}{{\cal D}^2}\,
\dot\alpha\, \dot\sigma^2 \nonumber\\
&& \qquad\qquad\qquad\qquad + \frac{20 p_L^6 + 96 p_L^4\, p_T^2 -
39 p_L^2\, p_T^4 - 34 p_T^6}{p^2\, {\cal D}^2}\, \dot\sigma^3
\Bigg\} \nonumber
\end{eqnarray}
\begin{eqnarray}
\Omega_{44}^2 &=& p^2 + \frac{\dot\phi^2 + \dot\phi_1^2}{4 M_p^2}
- \frac{1}{4}\, \dot\alpha^2 - \frac{p_L^2 - 2 p_T^2}{2 p^2}\,
\left( 4\, \dot\alpha - \frac{p_L^4 + 50 p_L^2\, p_T^2 - 5
p_T^4}{p^2\, \left( p_L^2 -2 p_T^2\right)}\, \dot\sigma \right)\,
\dot\sigma \nonumber\\
&& + \left[ V'(\phi) + 2\, \dot\phi\, \dot\alpha + 2\, \frac{p_L^2
- 2 p_T^2}{p^2}\, \dot\phi\, \dot\sigma \right]\,
\frac{f'(\phi)}{f(\phi)} - \dot\phi^2\,
\frac{f''(\phi)}{f(\phi)} \nonumber\\
&& - \frac{\tilde{p}_A^2\, p^2\, p_T^2}{2 M_p^2\, f(\phi)^2\,
{\cal D}^2}\, \Bigg[ \frac{4\, \tilde{p}_A^2}{M_p^2\, f(\phi)^2}\,
\frac{p_T^2}{p^2} + \frac{2 M_p^2\, {\cal D}^2}{p^2\, p_T^2}\,
\frac{f'(\phi)^2}{f(\phi)^2} + \frac{4\, \left( \dot\phi^2 +
\dot\phi_1^2 \right)}{M_p^2} - 20\,
\frac{p^2}{p_T^2}\, \dot\alpha^2 \nonumber\\
&& - 4\, \frac{10 p_L^4 + 17\, p_L^2\, p_T^2 + p_T^4}{p^2\,
p_T^2}\, \dot\alpha\, \dot\sigma - \frac{\left( 2 p_L^2 + 5 p_T^2
\right)\, \left( 10 p_L^4 - p_L^2\, p_T^2 - 5 p_T^4 \right)}{p^4\,
p_T^2}\, \dot\sigma^2 \Bigg] \label{Omega2}
\end{eqnarray}
where we have defined
\begin{equation}
{\cal D} \equiv 2 p^2\, \dot\alpha + \left( 2 p_L^2 - p_T^2
\right)\, \dot\sigma \label{calD}
\end{equation}
The canonically normalized modes $\{ V_1, \, V_2, \, H_{+}, \,
\Delta_{+} \}$ are related to the starting dynamical modes as
\begin{eqnarray}
V &=& e^{3\alpha/2}\, \left[ \delta\phi + \frac{p_T^2\,
\dot\phi}{{\cal D}}\, \Psi \right] \nonumber\\
V_1 &=& e^{3\alpha/2}\, \left[ \delta\phi_1 + \frac{p_T^2\,
\dot\phi_1}{{\cal D}}\, \Psi \right] \nonumber\\
H_{+} &=& e^{3\alpha/2}\, \frac{\sqrt{2}\, M_p\, p_T^2\, \left(
\dot\alpha + \dot\sigma \right)}{{\cal D}}\, \Psi \nonumber\\
\Delta_{+} &=& e^{3\alpha/2}\, \frac{p_T}{i\, p}\, \left[
\frac{\tilde{p}_A}{f(\phi)}\, \frac{p_T^2}{{\cal D}}\, \Psi +
e^{-\alpha + 2\sigma}\, f(\phi)\, \alpha_1 \right]
\label{defncan2dS}
\end{eqnarray}
When the universe becomes isotropic at the end of the first stage,
we have $\dot\sigma=0$ (we also rescaled the initial scale factors
such that $\sigma=0$ when the universe isotropizes). Using the
results of the equations
(\ref{pert-long-k}-\ref{long-2d-xi}-\ref{long-2d-metric}), the
canonical modes in the isotropic limit is related to the
perturbations defined in the longitudinal gauge as
\begin{equation}
h_{+} = - \frac{\sqrt{2}}{M_p}\, e^{-3\alpha/2}\, H_{+} \,\,\,\, ,
\,\,\, V = v = e^{3\alpha/2} \left( \delta\phi^{(L)} +
\frac{\dot\phi}{\dot\alpha}\, \Psi \right) \,\,\,\, , \,\,\, V_1 =
v_1 = e^{3\alpha/2} \left( \delta\phi_1^{(L)} +
\frac{\dot\phi_1}{\dot\alpha}\, \Psi \right) \label{Vv}
\end{equation}
where $v, \, v_1$ are the standard Mukhanov-Sasaki
variables~\cite{musa} (defined with respect to the action in the
cosmic time). All the above quantities are gauge invariant. These
relations will be useful when computing the curvature (${\cal R}$)
and $h_+$ gravitational wave polarization spectrum.

Note that the final action (\ref{action2dscan}) is of the generic
form (\ref{genericaction}), with the corresponding matrix ${\bf
X}$ given by
\begin{equation}
{\bf X} = \left( \begin{array}{ccc} 0 & 0 & {\cal F} \\
0 & 0 & {\cal G} \\ {\cal F} & {\cal G} & 0 \end{array} \right)
\end{equation}
In this case, the solution of the equation $\dot{\bf U} + i\, {\bf
X}\, {\bf U} = 0$ is more challenging, since the matrix that
diagonalizes ${\bf X}$ is given by
\begin{equation}
{\bf R} = \frac{1}{\sqrt{{\cal F}^2 + {\cal G}^2}}\, \left(
\begin{array}{ccc} {\cal G} & {\cal F}/\sqrt{2} & -{\cal
F}/\sqrt{2} \\ -{\cal F} & {\cal G}/\sqrt{2} & -{\cal G}/\sqrt{2}
\\ 0 & \sqrt{{\cal F}^2 + {\cal G}^2}/\sqrt{2} & \sqrt{{\cal F}^2 + {\cal G}^2}/\sqrt{2} \end{array} \right) \,\,\,\, ,
\,\,\, D_X = \left( \begin{array}{ccc} 0 & 0 & 0 \\ 0 &
\sqrt{{\cal F}^2 + {\cal G}^2} & 0 \\ 0 & 0 & -\sqrt{{\cal F}^2 +
{\cal G}^2} \end{array} \right)
\end{equation}
which is not a constant matrix, unlike the case for the $2d$
vector modes. However, the time derivatives of ${\bf R}$ are
suppressed by slow-roll parameters, therefore an approximate
solution for ${\bf U}$ can be found. We will discuss the details
and the computation of the power spectra of the curvature perturbation 
${\cal R}$ and $h_+$ in a separate publication.

\end{document}